\documentclass[a4paper,12pt]{article}
\pdfoutput=1
\usepackage{epsfig}
\usepackage{amssymb}
\usepackage{amsfonts}
\usepackage{amsmath}
\usepackage{mathtools}
\usepackage{euscript}
\usepackage{verbatim}
\usepackage{latexsym}
\usepackage{graphicx}
\usepackage{caption}
\usepackage{float}
\usepackage{subcaption}
\usepackage{bbm}
\usepackage[normalem]{ulem}
\usepackage{soul}
\usepackage{listings}
\usepackage{esint}
\usepackage{hyperref}

\usepackage{booktabs}

\newif\ifdtup

\usepackage{esint}
\catcode`\@=11

\@addtoreset{equation}{section}

\def\@normalsize{\@setsize\normalsize{15pt}\xiipt\@xiipt
\abovedisplayskip 14pt plus3pt minus3pt%
\belowdisplayskip \abovedisplayskip
\abovedisplayshortskip \z@ plus3pt%
\belowdisplayshortskip 7pt plus3.5pt minus0pt}

\def\small{\@setsize\small{13.6pt}\xipt\@xipt
\abovedisplayskip 13pt plus3pt minus3pt%
\belowdisplayskip \abovedisplayskip
\abovedisplayshortskip \z@ plus3pt%
\belowdisplayshortskip 7pt plus3.5pt minus0pt
\def\@listi{\parsep 4.5pt plus 2pt minus 1pt
     \itemsep \parsep
     \topsep 9pt plus 3pt minus 3pt}}

\everymath{\displaystyle}
\relax

\catcode`@=12

\catcode`\@=11

\def\section{\@startsection{section}{1}{\z@}{3.5ex plus 1ex minus
   .2ex}{2.3ex plus .2ex}{\large\bf}}

\def\SymBoxes#1#2#3#4{\newdimen\un@t \un@t#3%
\raisebox{#1}{\rule{#2\un@t}{#4}\hskip-#2\un@t
\@tempdimb\un@t \advance\@tempdimb by-#4\@tempcntb#2\relax%
\@whilenum{\@tempcntb>0}\do{
\rule{#4}{\un@t}\hskip\@tempdimb \advance\@tempcntb by\m@ne}%
\hskip-#2\un@t \rule[\un@t]{#2\un@t}{#4}%
\rule[\un@t]{#4}{#4}\hskip-#4
\rule{#4}{\un@t}}\hskip-#4}                

\begin{document}

\newcommand{\beq}{\begin{equation}}
\newcommand{\eeq}{\end{equation}}
\newcommand{\bea}{\begin{eqnarray}}
\newcommand{\eea}{\end{eqnarray}}
\newcommand{\beas}{\begin{eqnarray*}}
\newcommand{\eeas}{\end{eqnarray*}}
\newcommand{\defi}{\stackrel{\rm def}{=}}
\newcommand{\non}{\nonumber}
\newcommand{\bquo}{\begin{quote}}
\newcommand{\enqu}{\end{quote}}
\renewcommand{\(}{\begin{equation}}
\renewcommand{\)}{\end{equation}}
\def \eqn#1#2{\begin{equation}#2\label{#1}\end{equation}}

\def\e{\epsilon}
\def\IZ{{\mathbb Z}}
\def\IR{{\mathbb R}}
\def\IC{{\mathbb C}}
\def\IQ{{\mathbb Q}}
\def\de{\partial}
\def\Tr{ \hbox{\rm Tr}}
\def\H{ \hbox{\rm H}}
\def\HE{ \hbox{$\rm H^{even}$}}
\def\HO{ \hbox{$\rm H^{odd}$}}
\def\K{ \hbox{\rm K}}
\def\Im{ \hbox{\rm Im}}
\def\Ker{ \hbox{\rm Ker}}
\def\const{\hbox {\rm const.}}
\def\o{\over}
\def\im{\hbox{\rm Im}}
\def\re{\hbox{\rm Re}}
\def\bra{\langle}\def\ket{\rangle}
\def\Arg{\hbox {\rm Arg}}
\def\Re{\hbox {\rm Re}}
\def\Im{\hbox {\rm Im}}
\def\exo{\hbox {\rm exp}}
\def\diag{\hbox{\rm diag}}
\def\longvert{{\rule[-2mm]{0.1mm}{7mm}}\,}
\def\a{\alpha}
\def\dag{{}^{\dagger}}
\def\tq{{\widetilde q}}
\def\p{{}^{\prime}}
\def\W{W}
\def\N{{\cal N}}
\def\hsp{,\hspace{.7cm}}
\def\Tl{\Tilde}

\def\br{\nonumber}
\def\IZ{{\mathbb Z}}
\def\IR{{\mathbb R}}
\def\IC{{\mathbb C}}
\def\IQ{{\mathbb Q}}
\def\IP{{\mathbb P}}
\def \eqn#1#2{\begin{equation}#2\label{#1}\end{equation}}

\newcommand{\C}{\ensuremath{\mathbb C}}
\newcommand{\Z}{\ensuremath{\mathbb Z}}
\newcommand{\R}{\ensuremath{\mathbb R}}
\newcommand{\rp}{\ensuremath{\mathbb {RP}}}
\newcommand{\cp}{\ensuremath{\mathbb {CP}}}
\newcommand{\vac}{\ensuremath{|0\rangle}}
\newcommand{\vact}{\ensuremath{|00\rangle}                    }
\newcommand{\oc}{\ensuremath{\overline{c}}}
\newcommand{\psizero}{\psi_{0}}
\newcommand{\phizero}{\phi_{0}}
\newcommand{\hzero}{h_{0}}
\newcommand{\psiin}{\psi_{\rh}}
\newcommand{\phiin}{\phi_{\rh}}
\newcommand{\hin}{h_{\rh}}
\newcommand{\rh}{r_{h}}
\newcommand{\rb}{r_{b}}
\newcommand{\psibnd}{\psi_{0}^{b}}
\newcommand{\psibndp}{\psi_{1}^{b}}
\newcommand{\phibnd}{\phi_{0}^{b}}
\newcommand{\phibndp}{\phi_{1}^{b}}
\newcommand{\gbnd}{g_{0}^{b}}
\newcommand{\hbnd}{h_{0}^{b}}
\newcommand{\zh}{z_{h}}
\newcommand{\zb}{z_{b}}
\newcommand{\man}{\mathcal{M}}
\newcommand{\hbr}{\bar{h}}
\newcommand{\tbr}{\bar{t}}

\begin{titlepage}
\def\thefootnote{\fnsymbol{footnote}}

\begin{center}
{\bf {\large Holomorphic Factorization at the Quantum Horizon} \\  
\vspace{0.1in}
}
\end{center}

\bigskip
\begin{center}
Chethan Krishnan$^a$\footnote{\texttt{chethan.krishnan@gmail.com}}, \ Pradipta S. Pathak$^a$\footnote{\texttt{pradiptap@iisc.ac.in}}, \ 
\vspace{0.1in}

\end{center}

\renewcommand{\thefootnote}{\arabic{footnote}}

\begin{center}

$^a$ {Center for High Energy Physics,\\
Indian Institute of Science, Bangalore 560012, India}\\

\end{center}

\noindent
\begin{center} {\bf Abstract} \end{center}
We identify a horizon-skimming limit under which wave equations around large classes of black holes allow a determination of their low-lying (quasi-)degenerate normal modes. Building on our recent work, we use these ``quantum horizon" normal modes to study the thermodynamics of the parent black holes. A key observation is that the UV inputs (the location of the UV regulator, the number of species, and the cut-off in the angular Casimir quantum number) can all be combined into the freedom in a single real parameter. Remarkably, this parameter has an interpretation as the central charge of a holomorphically factorized 2D CFT, and choosing it to be the Kerr-CFT value reproduces the black hole's detailed thermodynamics from the statistical mechanics of normal modes. This perspective provides a heuristic understanding for why the Kerr-CFT central charge is related to the angular momentum of the black hole. The black holes we consider include Kerr-Newman in 3+1 dimensions and Cvetic-Youm in 4+1 dimensions (with all six charges), and they need not be BPS or extremal. Our results show that a refined version of the 't Hooftian quantum gas can be made fully consistent with the thermodynamics of very general black holes. This ``mechanical" approach to the central charge is not directly reliant on asymptotic symmetries in the extremal limit, where the black hole is often unstable. 



\vspace{1.6 cm}
\vfill

\end{titlepage}

\tableofcontents

\setcounter{footnote}{0}


\section{A Universal Statistical Physics for Black Holes}

Wave equations around black holes can exhibit surprising simplicity, even when the black hole metric is complicated. Perhaps the oldest known hint of this is the observation that the Kerr wave equation is separable \cite{Carter, Carter2}. This separability remains true \cite{Keeler} even for very general (Cvetic-Youm) black holes in string theory \cite{CveticYoum}. In fact, the radial parts of such separated wave equations often seem significantly simpler than they had any right to be. To get a sense of the simplicity, we invite the reader to take a look at the (complicated) metric and the (simple) wave equation in e.g., Appendix A and equation (36) of \cite{CveticLarsen}. 

In this paper, we will take the point of view that this simplicity is not merely a curiosity, but is indicative of something deeper. Wave equations around black holes capture bulk perturbations. In a holographic context, the background black hole together with these perturbations can be viewed as a bulk representation of the black hole microstate sliver in the CFT. These states should therefore be responsible for the entropy of the black hole in the microcanonical ensemble, indicating that these perturbations should implicitly be aware of black hole thermodynamics. However, there are some immediate obstructions to adopting this point of view. We explain these below and review a potential resolution that has been advocated in \cite{Vaibhav, Pradipta, Vaibhav2}. 

In AdS/CFT, the expectation is that light single-trace operators acting on a heavy CFT microstate are captured in the bulk dual by light (scalar and other) fields in the black hole geometry \cite{ElShowk}. While this is natural from the point of view of large-$N$ factorization around a thermal state, it raises a puzzle. One expects that the level-spacing of black hole microstates should be exponentially dense in $N$, to account for the Bekenstein-Hawking entropy. But light single trace operators and their normal modes around the AdS vacuum are not exponentially dense, and around a black hole with a smooth horizon the spectrum is quasi-normal (i.e., non-unitary). In \cite{Vaibhav, Pradipta, Vaibhav2} we observed a possible way out of this quandary, building on recent work in \cite{synth, RMT, Riemann, ArnabSuman}\footnote{See also the historically significant work in \cite{tHooft, SusskindThorlaciusUglum} as well as the more recent \cite{Souvik, Suchetan, Bobby, Suman}.}. It was noted that the correct (non-perturbative in $G_N$) degeneracy is reproduced by the normal modes of the bulk fluctuations, {\em if} there is a Planckian stretched horizon in the UV complete bulk description of a black hole microstate. A crucial point that was emphasized in \cite{Vaibhav} was that one should work {\em not} with the stretched horizon vacuum, but with the excitations at a sliver of energy at the mass of the black hole. Remarkably, it was found that the exterior correlators of this set up are indistinguishable (up to exponentially suppressed corrections) from those of a smooth horizon. This guarantees the emergence of the Hartle-Hawking correlator \cite{Vaibhav, Vaibhav2} in the bulk EFT description. It also provided a bulk Hilbert space implementation of a refined version of some old suggestions of 't Hooft \cite{tHooft, Solodukhin} and of Israel-Mukohyama \cite{Israel} in an AdS/CFT setting. Further, \cite{Vaibhav2} showed that the effective correlations in the exterior have a natural analytic continuation into the interior in the large-$N$ limit, providing a new perspective on the principle of equivalence at the horizon -- despite the manifest absence of smoothness in the UV complete description. In effect, this provides a reasonably precise way to view the stretched horizon as a finite-$N$ ingredient. A detailed matching of the entropy, temperature and the Hartle-Hawking correlator (together with the exponential entropic suppression of the variance of the correlator) were obtained from the quantum field theory with such a quantum horizon in \cite{Vaibhav, Pradipta, Vaibhav2}.

These observations strongly suggest that normal modes of black hole wave equations are approximate proxies for the UV complete spectrum, {\em responsible} for black hole thermodynamics. They also raise the possibility that the underlying reason for the surprising simplicity of black hole wave equations is their deep connection (outlined above) with states in the black hole microstate sliver. In other words, it is natural that the simplicity of black hole thermodynamics is reflected in some manner in the wave equations around black holes -- which is indeed what we noted in the opening paragraph.

This is promising, but there are two caveats to the results in \cite{Vaibhav, Pradipta}. The first is that the calculation needed three fairly heuristic UV inputs -- (a) the number $N_S$ of fields (species) excited at the stretched horizon\footnote{In \cite{Vaibhav, Pradipta} $N_S=1$ was implicit in the calculations because we were working with a single scalar. But the more general case is a trivial generalization. See the related discussions in \cite{Nomura}. Note that in counting $N_S$ we treat each spin degree of freedom as distinct. A slight further generalization is to allow distinct $l_{cut}$ and $\epsilon$ for each species. But the effective central charges simply add, and since only the total central charge is important for our discussion, we will not clutter our notation by incorporating this.}, (b) the location $\epsilon$ of the stretched horizon, and (c) a choice of cut-off $l_{cut}$ in the angular quantum number. The precise definitions of these quantities are not important at the moment; they will be elaborated upon in later sections. What matters for the present discussion is that there are multiple (superficially) distinct inputs needed for the calculations in \cite{Vaibhav, Vaibhav2}. It was appreciated in \cite{Vaibhav} that some of this freedom was superfluous, but a systematic discussion was not undertaken\footnote{The presentation of \cite{Vaibhav2} incorporates some of the lessons we have learnt in the present paper.}. In this paper, we will note that these heuristic UV inputs combine into the freedom in a {\em single} real parameter. We will further observe that this parameter has a direct interpretation as the central charge of a holomorphically factorized 2D CFT. The successful matching of the detailed thermodynamics done in \cite{Pradipta, Vaibhav} can be understood as arising from fixing the normal mode central charge correctly\footnote{The fact that the expressions are such that a matching is {\em possible} is non-trivial. This can ultimately be traced to a quasi-degeneracy of the normal modes in the  angular quantum number, and it was noted in \cite{Pradipta} that this was the underlying mechanism responsible for the area-scaling of entropy first noted by 't Hooft \cite{tHooft} -- without this, we would end up with the usual (Planckian) black body volume scaling. This angular degeneracy is a result of the redshift at the horizon, and does not arise if one simply puts a cut-off in flat space.}. The central charge perspective we will present in this paper was not manifest in the discussions of \cite{tHooft, Vaibhav, Pradipta}. We will see below that this perspective has the virtue that it naturally extends to more general black holes.


The second caveat in the results of \cite{Vaibhav, Pradipta} is that the discussion was limited to the BTZ black hole. Much of the present paper will be about generalizing these results to more general black holes. In particular, we will discuss Kerr-Newman black holes in 3+1 dimensions and the general black holes in string theory discovered by Cvetic and Youm \cite{CveticYoum} which are most directly discussed in 4+1 dimensions. Our black holes do not require extremality or supersymmetry, and are described by the mass, rotations and charges -- in the case of the Cvetic-Youm black holes we will have three $U(1)$ charges and two rotations on top of the mass. The key observation that makes these systems tractable is indeed the fact that their wave equations have remarkable properties. It has been noted before \cite{CveticLarsen, CMS, CK} that under a ``near-region" approximation, the wave equations on these black holes take a simple hypergeometric form. This form was used to associate certain dimensionless left and right temperature to these black holes in \cite{CMS, CK}. By postulating that the central charges of these black holes have the same form that they took in the extremal limit \cite{Guica, LMP}, the entropy and thermodynamics of the black holes were correctly reproduced via the Cardy formula \cite{CMS, CK}. We will refer to this collection of ideas as the (broadly defined) Kerr-CFT correspondence. 

In this paper, we will identify a variant of the limit considered in \cite{CMS, CK} and obtain closely related hypergeometric forms for the black hole wave equations. We call this the ``horizon-skimming limit". The forms of the wave equations that emerge retain the dimensionless left and right temperatures as well as the so-called hidden conformal symmetry. But they also have the key feature that the natural low-lying normal modes in this limit, carry the black hole entropy. The modes in the conventional near-region approximation \cite{CMS, CK} do not have this property. In higher dimensions, there are complications due to the Casimir quantum number being distinct from the azimuthal quantum numbers. But the horizon-skimming limit allows us to generalize our (rotating) BTZ discussion in \cite{Pradipta} to incorporate this in a natural way and determine the (quasi-)degenerate normal modes. Their statistical mechanics has the structure of a holomorphically factorized 2D CFT thanks to the hidden conformal symmetry, and the left and right temperatures together with the central charge is therefore sufficient to determine the thermodynamics completely.  Correctly reproducing the detailed thermodynamics of these black holes is equivalent to the statement that the central charge of the normal mode quantum gas is precisely the Kerr-CFT central charge.


Our result in this paper can be viewed as a refined version of 't Hooft's observation that the area-scaling of black hole entropy can be obtained from a brick wall cut-off. Some of the advantages of our refinement were listed in Section 2.1 of \cite{Vaibhav2} (see also \cite{Pradipta}). Here we see that the approach is in fact more general, and has the necessary structure to reproduce the correct thermodynamics of very general black holes. The UV inputs are encoded in a single real parameter which has the interpretation as the central charge of the normal mode quantum gas. The left and right moving temperatures were obtained from bulk wave equations in \cite{CMS, CK}. Their origins and significance as auxiliary wave equations seemed somewhat mysterious there, but the discussion here gives them a very physical and central role. The wave equations describe the dynamics of the low-lying normal modes, and the temperatures obtained in \cite{CMS} via a Rindler argument are very directly physical. It also gives a mutually self-consistent picture for the holomorphically factorized structure of wave equations vs. black hole thermodynamics -- the latter arises from the normal modes of the former.

A final noteworthy point we will mention here is that the Kerr-CFT central charges of black holes are known to be proportional to their angular momenta \cite{Guica, LMP}. As we will see, the central charges we find from our normal mode calculations are automatically the angular degeneracies associated to certain {\em total} angular quantum numbers ($l_{cut}$). Since the normal mode picture is supposed to give a ``mechanical" origin for the charges of black holes, this is encouraging. The central charges are proportional to the black hole angular momenta, because it is the angular momentum modes that are carrying the entropy. Equally interestingly, the codimension-2 (i.e., area) scaling of black hole entropy in Kerr-CFT comes from the fact that the central charge ($\propto$ angular momentum) has codimension-2 scaling. This is again completely automatic in our set up because the number of angular momentum modes with total angular momentum $l$ scales as $l^{d-2}$ in $d$ spacetime dimensions. 

Note that the BTZ case is an outlier from this perspective -- because the central charge $c=\frac{3L}{2G_3}$ in that case is not directly related to the angular momentum of the black hole. This is true for us, and this is true in Kerr-CFT. We discuss some hints regarding this distinction in Appendix \ref{holo}. The central charge in the more general cases is proportional to a {\em higher dimensional} angular momentum. In Appendix \ref{AdS2-sec}, we show that normal modes of extremal black holes with an AdS$_2$ near-horizon region do not have the crucial quasi-degeneracy required for the thermodynamics to work out correctly. This is a qualitative suggestion that extremal/BPS black holes are qualitatively different (at least in regards to normal modes) from more generic black holes. 

In the next section, we start with the holomorphically factorized wave equation that plays a crucial role in our discussion -- this is closely related to previous results \cite{CveticLarsen, CMS, CK}. We demonstrate these ideas for Kerr-Newman and Cvetic-Youm black holes in the subsequent sections. This involves some important distinctions from BTZ. Motivated by our horizon-skimming limit, we identify more general versions of the Kerr-CFT correspondence for 5D black holes than what has been discussed previously \cite{LMP, CK}. Various appendices deal with various technical asides -- some of which may be of conceptual interest. We review our main results and try to extract some general lessons in the final Discussion.




\section{Horizon-Skimming in BTZ} \label{BTZNormalModes}

We will start with the scalar field equation in the BTZ background and re-write it in a form that we will call the horizon-skimming wave equation. The motivation is that this structure has universal significance -- in later sections, we will see that both Kerr-Newman and Cvetic-Youm wave equations have connections to this form in a horizon-skimming limit (which is a variant of the limits considered in \cite{CMS}). Because the structure is universal, we will develop the statistical mechanics of this system here, and will find in later sections that it connects naturally with the thermodynamics of the underlying black holes in a unified way. 

We start by presenting the radial differential equation of a massless scalar in BTZ \cite{ArnabSuman, Pradipta}
\begin{equation} \label{RotatingBTZradialr}
    \frac{1}{r} \frac{d}{dr}\left(f^2 r \frac{d R(r)}{dr}\right) + \left(\frac{1}{f^2} \left(\omega - \frac{r_+ r_-}{r^2 L} m\right)^2 - \frac{m^2}{r^2}\right) R(r) = 0
\end{equation}
where $m$ is the $S^1$ quantum number and $f^2 = \frac{(r^2 - r^{2}_+)(r^2 - r^{2}_-)}{r^2 L^2}$, and $r_+$, $r_-$, and $L$ stand for the outer horizon, inner horizon, and AdS length, respectively. 
Using the dimensionless radial coordinate $z = \frac{r^2 - r^{2}_+}{r^2 - r^{2}_-}$, where inner and outer horizons correspond to $z \rightarrow -\infty$ and $z = 0$ respectively and $z = 1$ corresponds to AdS boundary, the differential equation \eqref{RotatingBTZradialr} becomes
\begin{equation} \label{RotatingBTZradialz}
    \frac{d}{dz}\left(z \frac{d}{dz} R(z)\right) + \frac{1}{4}\left[\frac{1}{z(1-z)}\left(\frac{\omega}{\kappa_+} - \frac{m \Omega^+_{\rm H}}{\kappa_+}\right)^2 - \frac{1}{1-z} \left(\frac{\omega}{\kappa_-} - \frac{m \Omega^-_{\rm H}}{\kappa_-}\right)^2\right]R(z) = 0
\end{equation}
Here $\kappa_\pm$ are the surface accelerations and $\Omega^\pm_{\rm H}$ are the rotational velocities at the outer and inner horizons, respectively. They are given as
\bea
& \kappa_\pm = \frac{r^{2}_+ - r^{2}_-}{r_\pm L^2} \\
& \Omega^\pm_{\rm H} = \frac{r_\mp}{r_\pm L}\\ \label{Omega_- BTZ}
& \frac{\Omega^-_{\rm H}}{\kappa_-} = \frac{\kappa_- \Omega^+_{\rm H}}{\kappa^2_+}
\eea
Since BTZ is locally AdS$_3$, \eqref{RotatingBTZradialz} naturally has an ${\rm SL}(2,\mathbb R)_L \times {\rm SL}(2,\mathbb R)_R$ symmetry. For this purpose, it is convenient to adopt ``conformal” coordinates $(w_\pm, y)$:
\bea \nonumber
& w^+  = \sqrt{z} \ e^{2 \pi T_R \phi + 2 K_R t},\\ \label{ConfCoordRotBTZ}
& w^- = \sqrt{z} \ e^{2 \pi T_L \phi + 2 K_L t},\\ \nonumber
& y = \sqrt{1-z} \ e^{\pi (T_R + T_L) \phi + (K_L + K_R) t},
\eea
where
\bea \label{CanTLRBTZ}
&  T_{L,R} = \frac{r_+ \pm r_-}{2 \pi L},\\ \label{KLR}
&  K_{L,R} = \mp \frac{r_+ \pm r_-}{2 L^2}.
\eea
This type of holomorphically factorized structure was first introduced in \cite{CMS} for the Kerr black hole. For BTZ, the result is natural because of the connection with AdS$_3$, see e.g. \cite{Kamali} who have formulas closely related to what we work with above. We will discuss some further aspects of these holomorphically factorized coordinates in Appendices. A crucial fact that will be important to us, which was explained in \cite{CMS}, is that the $T_L, T_R$ have interpretation as (dimensionless) left and right temperatures. 

For the moment, our goal is only to write the wave equation in terms of the new parameters $T_L, T_R, K_L$ and $K_R$. In terms of the ansatz $\Phi(t,z,\phi) = e^{-i \omega t + i m \phi} R(z)$ we find:
\bea \nonumber
& \partial_z (z \ \partial_z R(z)) + \frac{1}{4}\left[\frac{1}{z(1-z)} \left(\frac{T_L + T_R}{2 (K_R T_L - K_L T_R)} \omega + \frac{K_L + K_R}{2 \pi (K_R T_L - K_L T_R) }m\right)^2 - \right. \\ \label{CasimirRadialEqn}
& \qquad \left. \frac{1}{1-z} \left(\frac{T_L - T_R}{2 (K_R T_L - K_L T_R)} \omega + \frac{K_L - K_R}{2 \pi (K_R T_L - K_L T_R) }m\right)^2\right]R(z) = 0
\eea
This is simply \eqref{RotatingBTZradialz} given the expressions of $T_{L,R}$ and $K_{L,R}$ from \eqref{CanTLRBTZ} and \eqref{KLR}. 

In the BTZ case, the $T_L, T_R, K_L$ and $K_R$ are a redundant description of the parameters that describe the black hole, namely $r_+, r_-$ and $L$. The point however is that this form of the wave equation will arise universally in the horizon-skimming limits of higher dimensional black holes in the next sections. The difference will be in the definitions of $T_L, T_R, K_L$ and $K_R$ in terms of the parent black hole parameters.  Closely related (but not identical) radial equations were written down in \cite{CMS, CK} and various other papers in the context of the Kerr-CFT discussion -- we will clarify the connection to our approach in the next section.

In the rest of this section, our goal will be to write down the normal modes and their statistical mechanics for the horizon-skimming wave equation, while treating the $T_L, T_R, K_L$ and $K_R$ as the basic variables. Once we identify the horizon-skimming limits of the higher dimensional black holes in later sections, we will then be able to adapt the results here to the statistical mechanics of higher dimensional black holes.

\subsection{Normal Modes} \label{CasimirNormalModes}

We will write the general solution of \eqref{CasimirRadialEqn} as
\beq \label{CasimirRadialSoln} 
R(z) = e^{\pi \alpha} \ z^{-i \alpha} \ C_{2} \ {}_{2}F_{1}(a^*,b^*,c^*,z) + e^{-\pi \alpha} \ z^{i \alpha} \ C_{1} \ {}_{2}F_{1}(a, b, c, z)
\eeq
where $*$ denotes complex conjugate and $\alpha, a, b, c$ are given by
\bea \label{alphaabcCasimir}
  & \alpha = \frac{T_L + T_R}{4(K_R T_L - K_L T_R)} \Tl \omega\\
  & a = \frac{i \Tilde{\omega} T_R}{2 (K_R T_L - K_L T_R)}+ \frac{i m}{2 \pi (T_L + T_R)}\\
  & b = \frac{i \Tilde{\omega} T_L}{2 (K_R T_L - K_L T_R)}- \frac{i m}{2 \pi (T_L + T_R)}\\
  & c = 1 + 2 i \alpha
\eea
with  $\Tl \omega$ defined as \footnote{This definition makes $\Tl \omega$ the co-rotating mode. As we will see in the next sections, 
\bea
\Omega_{\rm H}^+= -\frac{K_L + K_R}{\pi (T_L + T_R)}
\eea
for the black holes we consider, where $\Omega_{\rm H}^+$ is the outer horizon angular velocity.}
\beq \label{tildeomegagen}
\Tl \omega \equiv \omega + m \frac{K_L + K_R}{\pi (T_L + T_R)}
\eeq 
The $z$-coordinate used in defining conformal coordinates \eqref{ConfCoordRotBTZ} for rotating BTZ has $z=1$ as the asymptotic boundary and $z=0$ as the black hole horizon. It turns out that for the asymptotically flat black holes in 3+1 and 4+1 dimensions that we will consider in the next sections also, the boundary\footnote{But note that the horizon-skimming form of the wave equation arises {\em after} a suitable limit in the full wave equation.} is at $z=1$ and the horizon is at $z=0$. The horizon-skimming limit leads to an effective AdS$_3$ box, and therefore it is natural to demand normalizability at boundary and Dirichlet near the horizon as done in \cite{Pradipta}.

With these observations, the calculation of the normal modes proceeds in parallel to that done for BTZ in \cite{RMT, Pradipta}. The discussion below only provides the main idea and glosses over some subtleties for which the reader should consult \cite{Pradipta}. In particular, we have considerable confidence in the quasi-degeneracy of the spectrum up to very high values of $m$, if the stretched horizon location ($\sim$ Planck length) is sufficiently small. This requires explicit numerical calculations to establish, even though the calculation of the partition function we present here is completely analytical.

Near $z \rightarrow 1$, the radial solution in \eqref{CasimirRadialSoln} looks like
\beq \label{NearBoundaryCasimirSoln }
   \frac{C_1 e^{-\pi \alpha} \Gamma (c)}{\Gamma (1+b) \Gamma (1+a)}+\frac{C_2 e^{\pi \alpha}\Gamma (c^*)}{\Gamma (1+b^*) \Gamma (1+a^*)}+\mathcal{O}\left(z-1\right)
\eeq
Demanding the analogue of normalizability relates $C_1$ and $C_2$ and the radial solution becomes
\begin{equation} \label{RadialSolnFinal Casimir}
   R(z) = C_{1} \ e^{-\pi \alpha} \left(z^{i \alpha} \ {}_{2}F_{1}(a,b,c,z) - z^{-i \alpha}\frac{\Gamma (c) \Gamma \left(1+a^*\right) \Gamma \left(1+b^*\right)}{\Gamma \left(c^*\right)\Gamma (1+a) \Gamma (1+b)}\ {}_{2}F_{1}(a^*, b^*, c^*, z)\right)
\end{equation}
By doing a near horizon expansion ($z \rightarrow 0$) of \eqref{RadialSolnFinal Casimir}, we get 
\begin{equation} \label{PhaseEqnCasimir}
    R(z) \approx C_{1}(T_{1} z^{-i \alpha} - z^{i\alpha})
\end{equation}
where,
\beq
   T_1 =  \frac{\Gamma (c) \Gamma \left(1+a^*\right) \Gamma \left(1+b^*\right)}{\Gamma \left(c^*\right)\Gamma (1+a) \Gamma (1+b)}
\eeq
In \eqref{PhaseEqnCasimir}, we demand a Dirichlet boundary condition at the stretched horizon radius $z = z_o$. The precise form of the boundary condition is in fact not important for our claims below, as long as they are non-dissipative.  The condition results in the phase equation
\begin{equation}
    T_1 = z^{2 i \alpha}_o,
\end{equation}
where we note that $T_1$ is manifestly a pure phase given the definitions of $a, b, c$. After some algebra, this can be turned into
\begin{align} \label{CasimirPhaseEqnCosine}
&\cos\left({\rm Arg}\left[\Gamma\left(\frac{i (T_L + T_R) \ \Tl \omega}{2(K_R T_L - K_L T_R)}\right)\right] + {\rm Arg}\left[\Gamma\left(-\frac{i \Tilde{\omega} T_L}{2(K_R T_L - K_L T_R)} + \frac{i m}{2 \pi (T_L + T_R)}\right)\right] \right. \nonumber \\
&\qquad\left. + {\rm Arg}\left[\Gamma\left(-\frac{i \Tilde{\omega} T_R}{2(K_R T_L - K_L T_R)} - \frac{i m}{2 \pi (T_L + T_R)}\right)\right] - \frac{(T_L + T_R) \ \Tl \omega}{2(K_R T_L - K_L T_R)} {\rm log}\left(\sqrt{z_o}\right)\right) = 0
\end{align}
which will be sufficient for us to make the claims that we need regarding the normal modes. 

\subsection{Quasi-Degenerate Spectrum}

It is possible to determine the normal modes analytically by using analytic approximations for arguments of Gamma functions in some regimes of $\Tl \omega$. The resulting solution was called the ``analytic low-lying spectrum" in \cite{Pradipta} and it has weak $m$-dependence at low $m$\footnote{We will never need the detailed expression for this form of the spectrum, but some of its qualitative features will be explained and used later. It should not be confused with the degenerate approximation for the spectrum we use below which ignores the (weak) $m$-dependence completely.}. Ignoring this weak $m$-dependence results in the (quasi-)degenerate spectrum which is all we will need in this paper:
\begin{equation} \label{CasimirSpectrum}
    \frac{\Tl \omega \ (T_L + T_R)}{2(K_R T_L - K_L T_R)} \approx \frac{n \pi}{{\rm log}\left(\frac{1}{\sqrt{z_o}}\right) }
\end{equation}
where $n \in \mathbb{Z}^{+}$.

In the expression above, $\tilde \omega \equiv \tilde \omega(n,m)$, which means we have a degeneracy in $m$. Let us make a few comments about this. In a strictly degenerate limit, we will need a cut-off in $m$ to make sure that the partition function and entropy are finite and well-defined. We believe this will have to be ultimately understood as a stringy/UV input\footnote{It may be possible that specifying the microcanonical sliver energy may result in foregoing the need for such an explicit cut-off. This speculation was made in \cite{Vaibhav2} and it is  a logical possibility, given the fact that the spectrum is weakly rising with $m$ and therefore naturally comes with a cut-off. But at large $m$ we only have numerical control on the spectrum. It is ultimately the second quantized spectrum that we are interested in when specifying the black hole mass \cite{Vaibhav}, and having only numerical control makes checking things challenging. So we have not been able to conclusively establish whether such a natural cut-off can be used to regulate the black hole entropy to the Bekenstein-Hawking value. In the present paper, we will simply view it as an explicit UV input. This is the possibility that we tend to lean towards at the moment.}. In the present paper, we will introduce it explicitly, and this $m_{cut}$ is one of the UV inputs that we discussed in the previous section. 

The analytic low-lying spectrum can reproduce the weak $m$-dependence at low-$m$, but it breaks down completely and diverges at a large  enough value of $m$, which we can call $m_{max}$. But it was noticed in the BTZ discussion of \cite{Vaibhav, Pradipta, Vaibhav2} that the modes that were needed for supporting the entropy were modes that were below this $m_{max}$ \cite{Vaibhav} by a small hierarchy. In this paper, we will see that despite the different scalings involved in higher dimensions, a similar result holds. Even though a somewhat technical fact, we feel this may be important. We will discuss this in more detail in Appendix \ref{AnalyticSpec}.


\subsection{Normal Mode Quantum Gas}
\label{CasimirPartitionFunction}

The left and right moving structure of the horizon-skimming wave equation naturally suggests that we work out the statistical mechanics of normal modes, in a holomorphically factorized manner. We will further assume that there are $N_S$ species of fields excited at the stretched horizon and they all have the same $m_{cut}$ and $z_o$. This can be easily relaxed with each species having its own $m_{cut}$ and $z_o$, but our conclusions do not fundamentally change (as discussed in the previous section). Note also that different species can have different spins, what we really mean by $N_S$ is the count of the propagating degrees of freedom. 

In doing the calculations below, it is useful to introduce the notation
\beq \label{DegCasimirSpectrum}
 \omega' \equiv \frac{\Tl \omega \ (T_L + T_R)}{2(K_R T_L - K_L T_R)}   \approx \frac{n \pi}{{\rm log}\left(\frac{1}{\sqrt{z_o}}\right)} \equiv n \omega'_o
\eeq
where the first equality defines a dimensionless spectrum $\omega'$ and the last equality defines\footnote{In \cite{Vaibhav2} it was noted that in the BTZ case, the dimensionful analogue of $\omega'_o$ is set by the scrambling timescale -- which is also the time it takes for an infaller to reach the stretched horizon from the AdS boundary. Here also it is clear that what $\omega'_o$ captures is a $\log N$ effect.} $\omega'_o$. Note that since we have dimensionless left and right temperatures, the energy also has to be dimensionless. Of course, this is not a unique choice of dimensionless $\Tl \omega$. We can multiply $\omega'$ by any dimensionless number and it will remain dimensionless. But as we will see, this freedom simply amounts to a redefinition of the central charge and therefore can be absorbed into the one-parameter freedom we have. 

A related point is that we could try to be a bit more systematic and work directly with the canonically conjugate energy associated to the (left and  right) temperature variables. We do this in Appendix \ref{Canonical} for completeness, but the conclusions remain intact. The ingredients that make the calculation successful (namely the linear dependence of the spectrum on one quantum number and the degeneracy in the other ones) are unaffected by these changes. 

Let us do the calculation first for the right movers ($m >0$).   The partition function of the right moving modes is simply
\bea \label{PartitionFunctionforCasimirR}
   \log Z_R &=& -\frac{N_S}{\omega'_o} \int_{0}^{\infty} d\omega' \sum_{m=1}^{m_{cut}} \log(1-e^{-\beta_R \omega'})\\
   \implies Z_R(\beta_R) & = & \exp\left[\frac{\pi^2 N_S m_{cut}}{6 \beta_R \omega'_o}\right]
\eea
Since $\omega'$ is dimensionless, $\beta_{R}$ in the partition function is also dimensionless. (But it is important to remember that at this stage, it is simply a variable that defines the ensemble and not directly related to the physical right temperature of the black hole introduced earlier.) The first line is simply the sum over all the $n$'s and $m$'s with the cut-off at $m_{cut}$ imposed. The $n$-sum has been converted into an integral, and this integral is exactly doable. Taking an inverse Laplace transform, we can get the density of states at a given energy $E_R$:
\begin{equation} \label{DensityofStatesR}
  g_R(E_R) = \frac{1}{2\pi i} \int_{\beta-i \infty} ^{\beta +i \infty} d\beta_R \exp\left[\frac{\pi^2 N_S m_{cut}}{6 \beta_R \omega'_o}+\beta_R E_R\right]
\end{equation}
where $E_R$ is the energy of the right modes. This can be evaluated by a saddle point approximation. The saddle is at\footnote{We will not introduce a new variable for the value of the variable $\beta_R$ at the saddle. We will simply call it $\beta_R$ from now on, and it is the physical right temperature.} 
\bea \label{betaRCasimir}
\beta_R = \sqrt{\frac{\pi^2 N_S m_{cut}}{6 \omega'_o E_R }}
\eea
The density of states and therefore the entropy at energy $E_R$ are given by
\bea \label{SaddlePotRCasimir}
   & g_{R}(E_R) \approx \exp\left[2 \sqrt{\frac{\pi^2 N_S E_R}{6 \omega'_o}m_{cut}}\right],\\ \label{SRCasimir}
   & S_R(E_R) = \log(g_R( E_R)) \approx 2 \sqrt{\frac{\pi^2 N_S E_R}{6 \omega'_o}m_{cut}}
\end{eqnarray}
From \eqref{betaRCasimir} and \eqref{SRCasimir}, we get a Cardy formula for $S_R$ as a function of $T_R$
\beq \label{SRTRCasimir}
S_{R} = \frac{\pi^2}{3} \left(\frac{N_S m_{cut}}{\omega'_o}\right) T_{R}.
\eeq
For the left-movers, things work out precisely analogously (with the summation on $m$ now going from $-m_{cut}$ to $-1$), and we get the same expressions with the label $R$ changed to $L$.  From \eqref{SRTRCasimir} and its left-handed counterpart the central charge can be read off as
\beq \label{centralchargeCasimir}
c_L = c_R = \frac{N_S m_{cut}}{\omega'_o}.
\eeq
It is immediate to check that if we set this normal mode central charge to be equal to the Brown-Henneaux central charge, $c_L=c_R=\frac{3L}{2 G_N^{(3)}}$, this reproduces the BTZ black hole thermodynamics in detail \cite{BTZStrominger}.

Let us make one comment. We have assumed above that the $m_{cut}$ and $N_S$ for the left movers and right movers are the same. This leads automatically to a matched central charge on the left and right sides. But if we relax this assumption, we can get the matching either as a demand (e.g., vanishing gravitational anomalies) or we will discover it upon comparing with the Brown-Henneaux central charge.

\section{Kerr-Newman} \label{KNNormalModes}

With parameters $M$, $J = Ma$ and $Q$, the Boyer-Lindquist form of the Kerr-Newman metric is
\begin{align} \label{KNBoyerLindquist}
& ds^2 = \frac{\rho^2}{\Delta} d\Tl{r}^2 - \frac{\Delta - a^2 \sin^{2}\theta}{\rho^2} dt^2 + \rho^{2} d\theta^2 + \frac{\sin^{2}\theta}{\rho^2}((\Tl{r}^2+a^2)^2 - \Delta a^2 \sin^{2}\theta)d\phi^2 \nonumber\\ 
& - \frac{2 a\sin^{2}\theta (\Tl{r}^2 + a^2 - \Delta)}{\rho^2}d\phi dt    
\end{align}
where $\Tl{r}_+$ and $\Tl{r}_-$ are outer and inner horizons respectively:
\begin{equation} \label{KNvariables}
\begin{split}
    &\Delta = \Tl{r}^2 + a^2 + Q^2- 2 G^{(4)}_N M \Tl{r}, \ \hspace{0.2cm}  \rho^{2} = \Tl{r}^2 + a^2 \cos^{2} \theta,\\
    &\Tl{r}_{+} + \Tl{r}_{-} = 2 G^{(4)}_N M, \ \hspace{0.2cm} \Tl{r}_{+} \Tl{r}_{-} = a^2 + Q^2.\\
\end{split}
\end{equation}
With $G^{(4)}_N$ the $4$-dimensional gravitational constant, $J = \frac{\Tl r_+ + \Tl r_-}{2 G^{(4)}_N} \sqrt{\Tl r_+ \Tl r_- - Q^2}$ is the physical angular momentum of the black hole. 

For a massless scalar, after choosing the ansatz $\Phi(t,\Tl{r},\theta,\phi) = e^{-i\omega t + i m \phi} R(\Tl{r}) \psi(\theta)$ the angular and radial differential equations are:
\begin{equation} \label{KNangular}
    \left(\frac{1}{\sin \theta} \partial_{\theta}(\sin \theta \partial_{\theta}) - \frac{m^2}{\sin^{2}\theta} + (\Tl{r}_{+} \Tl{r}_{-}-Q^2)\omega^{2} \cos^{2} \theta\right) \psi(\theta) = -\kappa_{l} \psi(\theta),\\
\end{equation}
\begin{align} \label{KNradial}
&\Bigg(\partial_{\Tl{r}}((\Tl{r}-\Tl{r}_+)(\Tl{r}-\Tl{r}_-) \partial_{\Tl{r}} + \frac{1}{4}\left[\frac{\Tl{r}_{+}-\Tl{r}_{-}}{\Tl{r}-\Tl{r}_{+}}\left(\frac{\omega}{\Tl{\kappa}_+} - \frac{m  \Tl{\Omega}^{+}_{\rm H}}{\Tl{\kappa}_+}\right)^2 - \frac{\Tl{r}_{+}-\Tl{r}_{-}}{\Tl{r}-\Tl{r}_{-}}\left(\frac{\omega}{\Tl{\kappa}_-} - \frac{m  \Tl{\Omega}^{-}_{\rm H}}{\Tl{\kappa}_-}\right)^2\right]+ \nonumber\\ 
& \hspace{0.5cm} + (\Tl{r}^2 + (\Tl{r}_{+}+\Tl{r}_{-})(\Tl{r}+\Tl{r}_{+}+\Tl{r}_{-})-Q^2) \omega^2 - \kappa_{l}\Bigg)R(\Tl{r}) = 0   
\end{align} 
We will have more to say about the separation constant $\kappa_l$ later. The $\Tl{\kappa}_\pm$ and $\Tl{\Omega}^\pm_{\rm H}$  are given as
\bea \label{kappaplusminusKN}
& \Tl{\kappa}_\pm = \frac{\Tl r_+ - \Tl r_-}{2(\Tl{r}_{\pm}(\Tl{r}_{+} + \Tl{r}_{-})-Q^2)} \\ \label{omegaplusminusKN}
& \Tl{\Omega}^{\pm}_{\rm H} = \frac{\sqrt{\Tl{r}_{+} \Tl{r}_{-}-Q^2}}{\Tl{r}_\pm (\Tl{r}_+ + \Tl{r}_-)-Q^2} \\ \label {ConstraintKN}
& \frac{\Tl{\Omega}^-_{\rm H}}{\Tl{\kappa}_-} = \frac{\Tl{\Omega}^+_{\rm H}}{\Tl{\kappa}_+}
\eea
Redefining the radial coordinate via
\begin{equation} \label{Rtoz KN}
    z = \frac{\Tl{r}-\Tl{r}_+}{\Tl{r}-\Tl{r}_-}
\end{equation}
the outer and inner horizons move to $z = 0$ and $z \rightarrow - \infty$ respectively and the asymptotically flat boundary goes to $z \rightarrow 1$. The radial wave equation \eqref{KNradial} written as
\begin{align} \label{KNradialz}
&\Bigg(\frac{d}{d z} \Big(z \frac{d}{d z}\Big)+ \frac{1}{4}\left[\frac{1}{z(1-z)}\left(\frac{\omega}{\Tl{\kappa}_+} - \frac{m  \Tl{\Omega}^{+}_{\rm H}}{\Tl{\kappa}_+}\right)^2- \frac{1}{1-z}\left(\frac{\omega}{\Tl{\kappa}_-} - \frac{m  \Tl{\Omega}^{-}_{\rm H}}{\Tl{\kappa}_-}\right)^2\right]+ \\ 
& + \frac{1}{(1-z)^2}\Bigg(\Bigg(\Big(\frac{\Tl{r}_+ - \Tl{r}_- z}{1-z}\Big)^2 + (\Tl{r}_{+}+\Tl{r}_{-})\Bigg(\Big(\frac{\Tl{r}_+ - \Tl{r}_- z}{1-z}\Big)+\Tl{r}_{+}+\Tl{r}_{-}\Bigg)-Q^2\Bigg) \omega^2 - \kappa_{l}\Bigg)R(z) = 0.  \nonumber
\end{align}
is too complicated to be analytically solvable in a useful way. 

\subsection{Horizon-Skimming Limit}\label{kerrskimm}

We will define a ``horizon-skimming" limit for \eqref{KNradialz} via \footnote{It should be kept in mind that the energies of these modes are measured asymptotically, so one has to be careful in interpreting statements like ``their wavelength is longer than the size of the black hole".}
\bea \label{KNhorskimr}
\Tl r \sim \Tl r_+ \ll \frac{1}{\Tl \omega}
\eea
which can be viewed in $z$-coordinates as the statement that
\bea
\label{KNhorskimz}
 z \rightarrow 0 \ \ {\rm and} \ \ \Tl \omega\ \Tl r_+ \rightarrow 0.
\eea
Note that $\Tl r_+$ sets the biggest scale among the black hole parameters, therefore all other similar dimensionless quantities constructed from $\tilde \omega$ also go to zero in this limit.
The co-rotating modes $\Tl{\omega}$ in this geometry are defined by
\begin{equation} \label{ShiftedModesKN}
    \Tilde{\omega}(n,m) \equiv \omega(n,m) - m\Tl{\Omega}^+_{\rm H} = \omega(n,m) - m\frac{\sqrt{\Tl{r}_+ \Tl{r}_- - Q^2}}{\Tl{r}_+ (\Tl{r}_+ + \Tl{r}_-)-Q^2}.
\end{equation}
We wish to compute normal modes for $\tilde \omega$. 
Of course, to make the second approximation in \eqref{KNhorskimz} self-consistently, we need at least a sector of the normal modes to satisfy 
\beq
\Tl \omega \ll \frac{1}{r_+}
\eeq
A crucial observation is that the $\log$ term in the denominator of the low-lying normal modes \eqref{CasimirSpectrum} will provide this hierarchy ($\sim \log N$). Indeed the entire set of quasi-degenerate modes below $m_{cut}$ satisfy this condition for low values of $n$. This was crucial for the entropy/temperature discussion in Section \ref{CasimirPartitionFunction} to work.

The horizon-skimming limit is different from the {\em near-region} limit that was considered in \cite{CMS} (see also \cite{CK} for the 5D discussion). There the low modes considered were in $\omega$, not in $\tilde \omega$ -- i.e., they were not co-rotating. The rationale behind this choice was that this way, all terms except the $\kappa_l$ term in the second line of the radial equation \eqref{KNradialz} would drop. Furthermore, the \eqref{KNangular} becomes the usual spherical harmonic equation and therefore the separation constants $\kappa_l$ were taken to be $l(l+1)$. These are convenient technical demands that simplify the equations. No self-consistency check was available for the validity of the approximation, since the modes were declared to exist by fiat, presumably to be viewed as long wavelength {\em classical} waves near the horizon. At the quantum level, one must worry about (a) boundary conditions for the modes\footnote{Note that quasinormal boundary conditions lead to complex modes, and their scale is the same order as the horizon size.}, and (b) whether the modes one obtains after imposing the boundary condition can self-consistently satisfy the near-region approximation. 

This is a somewhat unsatisfying state of affairs. Since the goal of \cite{CMS} was to try and motivate the {\em quantum} origin of black hole entropy via the wave equation, it would be natural to view the modes as quantum. But what boundary condition should we use? Do the modes we obtain from the boundary conditions self-consistently satisfy a near-region type approximation? Can the black hole thermodynamics be understood in terms of them? 

In this paper (among other things) we answer all three of these questions -- the latter two in the affirmative -- by working with a horizon-skimming limit instead of a near-region limit\footnote{A limit similar to \eqref{KNhorskimr} was considered in equation B.1 of \cite{Pradipta}, but with $\omega$ instead of ${\Tl \omega}$. We only considered the $m=0$ modes for detailed calculations there, and in that sector $\omega =\tilde \omega$. In other words, horizon-skimming limit and near-region limit were largely identical in the limited context studied in \cite{Pradipta}. For a non-zero $m$, it is the $\Tl \omega$ normal modes that satisfy \eqref{KNhorskimr} and  exhibit the quasi-degeneracy in $m$.}. The idea is to consider a stretched horizon boundary condition for the scalar modes and compute the co-rotating normal modes in the horizon-skimming limit. We will find that the structures we found for the quasi-degenerate modes in the BTZ case are largely retained in the higher dimensional cases as well. The extra terms that are present in the second line of \eqref{KNradialz} have a natural and fairly innocuous understanding in this limit. To see this, we use \eqref{KNhorskimz} and \eqref{ShiftedModesKN} (which in turn implies $\omega \sim m \Tl \Omega^{+}_{\rm H}$) in the term containing $1/(1-z)^2$ in \eqref{KNradialz}. The radial equation \eqref{KNradialz} almost turns into the BTZ horizon skimming equation in this limit, but with an extra term from the second line of \eqref{KNradialz}. The latter takes the form
\beq \label{SpecialmodesKN}
\Big(m^2 (\Tl \Omega^+_{\rm H})^2 (\Tl{r}^2_+ + (\Tl{r}_{+}+\Tl{r}_{-})(2\Tl{r}_{+}+\Tl{r}_{-})-Q^2) - \kappa_l\Big) R(z). 
\eeq
Since there is a slight possibility of confusion, let us emphasize that the $m$ above is the azimuthal quantum number, and not the mass. Note that the separation constant $\kappa_l$ is only known numerically \cite{Teukolsky} since the term $\propto \omega^2$  in \eqref{KNangular} does {\em not} go to zero in the horizon-skimming limit. The expression multiplying $m^2$ is an $\mathcal{O}(1)$ number -- we make this precise in Appendix \ref{kappaextreme}.

To identify the normal modes, we need to solve the system with this extra term in the horizon skimming wave equation. While the wave equation in this limit is solvable in terms of hypergeometric functions, imposing the boundary conditions is difficult when the extra term \eqref{SpecialmodesKN} is present -- it goes into certain Gamma functions that make the normal mode phase equation unsolvable, analytically\footnote{The trouble is that we have not been able to extract complex phases of gamma functions whose arguments contain generic real and imaginary parts (other than numerically). The cases where we have succeeded in solving them using analytic methods usually contain special values for the real part.}. This is a new feature that makes higher dimensional black holes more involved than BTZ. 

We will now argue that the two key properties that allowed us to match the thermodynamics of the BTZ normal modes are still present in the equation, despite the presence of the extra term above. These key properties are (a) that there is a quasi-degeneracy in the spectrum in the angular quantum number directions, and (b) that we can explain the entropy by retaining only modes up to some angular momentum cut-off $l_{cut}$ within this quasi-degenerate regime. The way we will establish this, is by arguing that the extra term in \eqref{SpecialmodesKN} does not affect  the magnitude of the normal modes, nor the fact that there is an approximate degeneracy\footnote{A minor caveat in this statement is discussed in Appendix \ref{kappaextreme}.}. We have not been able to analytically prove this statement for arbitrary (allowed) choices of $m$ and $l$. But we can find strong evidence for it by solving it in two extreme cases. The first case is when 
\bea
m^2 (\Tl \Omega^+_{\rm H})^2 \Big(\Tl{r}^2_+ + (\Tl{r}_{+}+\Tl{r}_{-})(2\Tl{r}_{+}+\Tl{r}_{-})-Q^2\Big) \sim \kappa_l \label{case1}
\eea
so that \eqref{SpecialmodesKN} is approximately vanishing, and the second case is when $m=0$. The former case can be translated to  the horizon-skimming equations we solved in the previous section in terms of $K_L, K_R, T_R$ and $T_L$, and the latter case was solved in \cite{Pradipta}.

Before discussing these solutions and using the resulting modes to compute their statistical mechanics, we briefly pause to clarify why solving the system in these two cases is good evidence that the result is valid in the general case as well. The main point is that for any given value of $l$, the range of $m$ is bounded between $-l$ and $l$. It is straightforward to check that the function of ${\tilde r_+}$, ${\tilde r_+}$ and $Q$ that multiplies $m^2$ in the above expression is an $\mathcal{O}(1)$ number\footnote{Again, see a slightly more detailed discussion in Appendix \ref{kappaextreme}.} . This means that the term \eqref{SpecialmodesKN} is at most of $\mathcal{O}(l(l+1))$. The two cases we are considering therefore are roughly the two extreme cases (i.e, $m=0$ and $|m| \sim l$), and since the function involved is continuous we expect the resulting modes also to vary continuously. By showing that the modes are identical and degenerate to a first approximation in both extremes, we can therefore argue that they behave likewise in the intermediate cases as well. What is more, we have numerically checked this (quasi-)degeneracy very explicitly for various numerical values of $\tilde r_+, \tilde r_-$ and $Q$.

With this preamble, when \eqref{case1} holds, we can write \eqref{KNradialz} in the form \eqref{CasimirRadialEqn} after identifying \cite{CMS, WangLiu}
\bea  \label{TLRKerN}
& T_{L,R} = \left(\frac{\Tl \kappa_- + \Tl \kappa_+}{\Tl \kappa_- \mp \Tl \kappa_+}\right) \frac{\Tl \kappa_+}{2\pi \Tl \Omega^+_{\rm H}}\\ \label{KLRKerN}
&K_R = 0 \ , \ K_L = \frac{\Tl \kappa_- \Tl \kappa_+}{\Tl \kappa_+ - \Tl \kappa_-}.
\eea 
This allows us to straightaway write down the degenerate in $m$ low-lying modes from \eqref{CasimirSpectrum} as
\beq \label{KNSpectrum}
\frac{\Tl \omega}{\Tl \kappa_+} = \frac{n \pi}{\log \left[\frac{1}{\sqrt{z_o}}\right]}.
\eeq

The stretched horizon location $z_o$ is a free parameter at the moment, but if we follow 't Hooft, we should set it to be equal to a Planck length from the horizon. This philosophy has two drawbacks. One is that when the black hole is {\em not} spherically symmetric, a fixed geodesic length in the radial $z$-direction translates to a $z_0$ that is angle-dependent. A second problem is that in holographic theories, it is better to anchor the distance to the asymptotic boundary, and not the horizon \cite{Vaibhav2}. We will therefore simply view $z_0$ as part of the UV data that needs to be specified. It captures\footnote{Note that the $z$-coordinate is dimensionless.} a ``Planckian" or stringy length scale, but it is not necessarily {\em defined} by the idea that it is {\em exactly} a Planck length from the horizon. In fact (as we discussed), all of the UV data that goes into the calculation, results in an overall one-parameter freedom which has a natural interpretation as the total central charge. Even though we will not use it too much, it is useful to note that if we demand that the stretched horizon is at a Planck length, $z_o = \frac{l^2_p}{4 (\Tl r^2_+ + a^2)}$ at $\theta = 0$ (North Pole). 

Now we turn to the $m=0$ case, for which we computed the low-lying spectrum for the uncharged Kerr case in Appendix B of \cite{Pradipta}. We solved the radial differential equation \eqref{KNradial} using the horizon skimming limit \eqref{KNhorskimr} with $Q=0$. Radial solutions there can be trivially extended for a non-zero $Q$ \footnote{$Q$ finds its presence in $A$ and $B$ defined in equation B.11 of \cite{Pradipta}.}. This will again give us a quasi-degenerate spectrum, this time directly in $l$, see Figure 22 of \cite{Pradipta}\footnote{Figure 22 is for a fairly large value of $z_o$. We have checked that for smaller values, the degeneracy approximation is significantly better.}. We get the low-lying spectrum from equation B.23 after using the low $\omega$-approximation of Gamma functions in equation B.29 and B.30 
\beq \label{KNSpecmzero}
\frac{\omega}{\Tl \kappa_+} = \frac{n\pi}{\log\left[\frac{1}{ \ \sqrt{z_o}}\right]}
\eeq
Note that the two results are identical. Thus the two extreme cases $m = 0$ and $m \sim \sqrt {\kappa_l}$ (as well as the numerical interpolation) motivate the claim that the dimensionless degenerate spectrum, given by 
\beq 
\omega' = \frac{n\pi}{\log\left(\frac{1}{\sqrt{z_o}}\right)} \equiv n \omega'_o
\eeq
is naturally valid for other choices of $m$ as well at a given $l$, as long as $l$ is below some cut-off $l_{cut}$. The value of $l_{cut}$ is another piece of UV data.

\subsection{Partition Function and Thermodynamics} \label{KNPartitionFunction}

The calculation proceeds parallel to that in Section \ref{CasimirPartitionFunction} but with some physically significant differences in the counting of the angular momentum modes.

Firstly, since this is a $3+1$ black hole, the states will be labeled by $n,l,m$. Secondly, for any form of $\kappa_l$, we have $-l\leq m \leq l$. This means that we need to integrate over $m$ within this range. Finally, there will be an integration over $l$-modes with $2l$ degeneracy (coming from $m$-integral) from $0$ to $l_{cut}$. This will be done together with the integration along $\omega'$ captured by $n$.  For the right movers ($m>0$),  
\bea \label{PartitionFunctionKNR}
    \log Z_R &=& -\frac{N_S}{\omega'_o} \int_{0}^{l_{cut}} dl \int_{0}^{l} dm \int_{0}^{\infty} d\omega' \log(1-e^{-\beta_R \omega'})\\
   \implies Z_R(\beta_R) & = & \exp\left[\frac{\pi^2 N_S l^{2}_{cut}}{12 \beta_R \omega'_o}\right]
\eea
For the left-movers, things work out precisely analogously -- with the only difference being the integral on $m$ running from $-l$ to $0$.  The end result is similar to \eqref{SRTRCasimir}, but with a different form for the normal mode central charge: 
\beq \label{SLRTLRKN}
S_{L,R} = \frac{\pi^2}{3} \left(\frac{N_S l^{2}_{cut}}{2\omega'_o}\right) T_{L,R}
\eeq
The left and right temperatures are defined in \eqref{TLRKerN}. Identifying this with the Kerr-CFT central charge for this black hole \cite{WangLiu}
\beq \label{centralchargeKN}
c_L = c_R = \frac{N_S l^2_{cut}}{2 \omega'_o} = 12 J
\eeq
is guaranteed to reproduce the correct left and right (and therefore total) entropy:
\beq \label{SLRSimpleKN}
S_{L,R} = 2\pi \left(\frac{\Tl \kappa_- + \Tl \kappa_+}{\Tl \kappa_- \mp \Tl \kappa_+}\right) \frac{J \Tl \kappa_+}{\Tl \Omega^+_{\rm H}}.
\eeq
These expressions can be converted to the standard Kerr-Newman area formulas in (say) \cite{BinChen} using the formulas we presented earlier.

Let us re-iterate a couple of facts. Firstly, our calculation shows that the Kerr-Newman thermodynamics can be made consistent with a refined 't Hooftian statistical mechanics. The holomorphic factorization structure of the wave equation was crucial for this. Conversely, it is not clear how one would set up a satisfactory 't Hooftian calculation {\em without} exploiting that structure. We comment further about it in the Discussion. At the same time, determining the central charge from first principles is clearly an open challenge. Secondly, the calculation shows (as promised) that {\em all} the UV inputs can be combined into the freedom in a single real parameter -- with the interpretation as the central charge. It is also intriguing that the central charge in Kerr-CFT is proportional to the angular momentum of the black hole, while the normal mode central charge naturally relies on the angular momentum modes excited in the geometry.  As noted in the introduction the ``area" scaling of $J$ is captured by the power of $l_{cut}$ in the normal mode central charge. The matching of the scaling continues to hold in higher dimensions.


\section{Cvetic-Youm} \label{CveticYoum}

The most general black holes of the low energy effective action for type II/heterotic string theory, toroidally compactified down to 5 dimensions \cite{CveticYoum} are characterized by mass ($M$), two independent angular momenta ($J_L$ and $J_R$) and three $U(1)$ charges ($Q_1$, $Q_2$ and $Q_3$). These quantities are usually specified as \cite{CveticLarsen} 
\bea
& M = \frac{\mu}{2} \sum_{i=1}^{3} \cosh{2\delta_i}\\ \label{M5D}
& Q_i = \frac{\mu}{2} \sum_{i=1}^{3} \sinh{2\delta_i} \ ; \ i = 1,2,3\\ \label{Q5D}
&J_{R,L} = \frac{\mu}{2} (l_1 \pm l_2) (\prod_{i=1}^{3} \cosh{\delta_i} \mp \prod_{i=1}^{3} \sinh{\delta_i}) \label{JRL5D}
\eea
where $\mu$ is the mass parameter, $\delta_i$'s are ``boosts" and $l_1$ and $l_2$ are parameters associated to rotation. $\mu$ and $l_{1,2}$ in terms of the inner ($\bar r_-$) and outer horizon ($\bar r_+$) radius and angular momentum along $\phi$ and $\psi$ directions (see footnote 3 of \cite{CveticLarsen}) are given as 
\bea \label{mul125D}
   & \mu = \frac{(\bar r^2_+ + l^2_1)(\bar r^2_+ + l^2_2)}{\bar r^2_+} , \ \bar r_+ \bar r_- = l_1 l_2 \\ \label{Jphipsi5D}
   & J_{\phi,\psi} = J_R \pm J_L
\eea
The remarkable fact is that despite the complexity of the metric, the radial and angular wave equation in the geometry reduce to a surprisingly simple form \cite{CveticLarsen}:
\begin{align} \label{5dWaveEquation}
&\frac{\partial}{\partial x} \left[\left(x^2-\frac{1}{4}\right) \frac{\partial}{\partial x} R(x)\right]+\frac{1}{4}\left[\frac{1}{x-\frac{1}{2}}\left(\frac{\omega}{\Bar{\kappa}_+} - \frac{m_R \Omega_R}{\Bar{\kappa}_+} - \frac{m_L \Omega_L}{\Bar{\kappa}_+}\right)^2 \right. \nonumber \\
& \qquad \left. - \frac{1}{x+\frac{1}{2}}\left(\frac{\omega}{\Bar{\kappa}_-} - \frac{m_R \Omega_R}{\Bar{\kappa}_+} + \frac{m_L \Omega_L}{\Bar{\kappa}_+}\right)^2 + (x\Delta + M) \omega^2 - \Lambda \right]R(x) = 0.\\ 
\label{5dAngularEquation}
& \Bigg(\frac{1}{\sin 2\theta} \frac{\partial}{\partial\theta}\left(\sin 2\theta \frac{\partial}{\partial \theta}\right) - \frac{m_R + m_L}{\sin^{2}\theta} - \frac{m_R - m_L}{\cos^{2}\theta} + (l^{2}_1 + l^{2}_2)\omega^{2} + (l^{2}_2 - l^{2}_1)\omega^{2} \cos 2\theta\Bigg)\chi(\theta) = -\Lambda \chi(\theta) 
\end{align}
where we have separated the scalar as
\begin{equation} \label{probe ansatz}
   \Phi = R (r) \chi(\theta) e^{-i\omega t + i m_\phi \phi + i m_\psi \psi}
\end{equation}
$\Delta$ is a constant that can be expressed in terms of entropy ($S$) and temperature ($\beta_{H}^{-1}$) as $\Delta = \beta^{-1}_H S = \bar r^{2}_+ - \bar r^{2}_-$, $M$ is the mass of the black hole, $\Lambda$ is the eigenvalue of the angular Laplacian and $x$ is a radial coordinate, related to conventional Boyer-Lindquist-like coordinate $r$ that ranges from $0$ to $\infty$, by
\begin{equation} \label{xdef}
    x = \frac{\bar r^2 - \frac{1}{2}(\Bar{r}^{2}_+ + \Bar{r}^{2}_-)}{\Bar{r}^{2}_+ - \Bar{r}^{2}_-}
\end{equation}
In this coordinate, the outer and inner horizons at $\bar r_\pm$ are at $x = \frac{1}{2}$ and $x = -\frac{1}{2}$ respectively; and asymptotically flat region is at $x=\infty$. Left and Right angular velocities $\Omega_{L,R}$ and surface accelerations at outer and inner horizons $\Bar{\kappa}_{\pm}$ respectively, are defined in eqn $(17-18) \ {\rm and} \ 30$ of \cite{CveticLarsen}:
\bea \label{OmegaLR}
& \beta_H \Omega_{L,R} = \frac{2 \pi J_{L,R}}{\sqrt{\frac{1}{4}\mu^3 \left(\prod_i \cosh \delta_i \pm \prod_i \sinh \delta_i\right)^2 - J^{2}_{L,R}}}\\ \nonumber
& \frac{1}{\Bar{\kappa}_\pm} = \frac{\frac{1}{4} \mu^2 \left(\prod_i \cosh^{2} \delta_i - \prod_i \sinh^{2} \delta_i\right)}{\sqrt{\frac{1}{4}\mu^3 \left(\prod_i \cosh \delta_i - \prod_i \sinh \delta_i\right)^2 - J^{2}_R}} \pm \frac{\frac{1}{4} \mu^2 \left(\prod_i \cosh^{2} \delta_i - \prod_i \sinh^{2} \delta_i\right)}{\sqrt{\frac{1}{4}\mu^3 \left(\prod_i \cosh \delta_i + \prod_i \sinh \delta_i\right)^2 - J^{2}_L}} \\ \label{kappaplusminus}
\eea
The left and right excitations ($m_R$ and $m_L$) are related to the excitations along two azimuthal  circles of rotation, $\phi$ and $\psi$, each having a periodicity of $2\pi$, as $m_{\phi,\psi} = m_R \pm m_L$. 

Now, we will make a coordinate transformation 
\begin{equation} \label{xtoz5D}
    x = \frac{1}{2}\left(\frac{1+z}{1-z}\right)
\end{equation}
where, outer and inner horizons are given by $z = 0$ and $z \rightarrow -\infty$ respectively and asymptotically flat region corresponds to $z \rightarrow 1$. Then the radial wave equation \eqref{5dWaveEquation} written in the $z$ radial coordinate 
\begin{align} \label{WaveEqn5Dz}
&\frac{d}{d z} \left[z \frac{d}{d z} R(z)\right]+\frac{1}{4}\left[\frac{1}{z(1-z)}\left(\frac{\omega}{\Bar{\kappa}_+} - \frac{m_R \Omega_R + m_L \Omega_L}{\Bar{\kappa}_+}\right)^2 \right. \nonumber\\
&\left. - \frac{1}{1-z}\left(\frac{\omega}{\Bar{\kappa}_-} - \frac{m_R \Omega_R - m_L \Omega_L}{\Bar{\kappa}_+}\right)^2 + \frac{1}{(1-z)^2}\left(\left(\frac{1}{2}\left(\frac{1+z}{1-z}\right)\Delta + M\right)\omega^2 - \Lambda \right)\right] R(z) = 0 
\end{align}
is our starting point.

\subsection{Horizon-Skimming Limit}

The horizon-skimming limit is again given via 
\beq \label{5Dhorskimx}
\bar r \sim \bar r_+ \ll \frac{1}{\Tilde{\omega}} \ \ \ {\rm or} \ \ \
 z \rightarrow 0, \tilde \omega\ \bar r_+ \rightarrow 0.
\eeq
where the co-moving modes $\Tl \omega$ in this geometry can be taken to be 
\begin{equation} \label{5DShiftedModes}
    \Tilde{\omega}(n,m) \equiv \omega(n,m) - (m_L\Omega_L + m_R\Omega_R).
\end{equation}
This means that can use $z \sim 0$ and $\omega \sim m_L \Omega_{L} + m_R \Omega_{R}$ in the parenthesis multiplying $1/(1-z)^2$ in \eqref{WaveEqn5Dz}. Analogous to the Kerr-Newman case, the two extreme cases we will consider will be $m_L=m_R=0$ and 
\beq \label{Specialmodes5D}
(m_L\Omega_L + m_R\Omega_R)^2\left(\frac{\Delta}{2} + M\right) \sim \Lambda 
\eeq 
We will show that the spectrum is identical and (quasi-)degenerate in these two cases to extract the normal modes of the full system in the horizon-skimming limit. 
As before, note that the separation constant $\Lambda$ is only known numerically because $\omega$'s in \eqref{5dAngularEquation} are non-zero, but we expect it to be $\mathcal{O}(l(l+2))$.

\subsection{Myers-Perry} \label{MPmodes}

The higher dimensional wave equations we have presented above contain two azimuthal angles (and therefore two azimuthal quantum numbers). Our discussion of the horizon-skimming wave equations in the previous sections on the other hand, naturally came with only one azimuthal angle. In this paper, we will show that {\em any} general direction in the plane of the two angles can serve as a suitable choice for our purposes -- in other words, any $a\partial_\phi+b \partial_\psi$ will lead to a well-defined left and right temperature as well as left-right symmetric central charge. This is natural from our holomorphically factorized normal mode perspective, but historically it seems that only certain special cases have been considered in the literature. In this section, we will reproduce these old results. The central charge we find here has been obtained also from an asymptotic symmetry argument in the extremal limit by \cite{LMP}. 

The case with all the $U(1)$ charges turned off ($\delta_i = 0$) with $l_1$ and $l_2$ being independent is usually referred to as a doubly spinning Myers-Perry Black hole. In this black hole, the $\psi$-singlet, i.e., $m_\psi = 0$ wave equation has already been discussed in \cite{CK}. There the goal was to find evidence for Kerr-CFT in the higher dimensional case. Here, we will find out the normal modes for this system.  The $\psi$-singlet modes mean that our choice of $m$ is defined by $m_R = m_L \equiv m$. We will first look at the extreme case \eqref{Specialmodes5D} which loosely implies  $m \sim \sqrt{\Lambda}$. The equation then reduces to the horizon-skimming equations of Section \ref{BTZNormalModes}. 
Identifying \cite{CK}
\bea \label{TLRMP}
& T_{L,R} = \frac{1}{\pi}\frac{\bar \kappa_+ (\bar \kappa_- \pm \bar \kappa_+)}{(\bar \kappa_- - \bar \kappa_+)\bar\Omega_R - (\bar \kappa_- + \bar \kappa_+)\bar\Omega_L}\\ \label{KLR5D}
& K_{L,R} = \frac{\Omega_{R,L}\bar \kappa_+ \bar \kappa_-}{(\bar \kappa_- - \bar \kappa_+)\bar\Omega_R - (\bar \kappa_- + \bar \kappa_+)\bar\Omega_L}
\eea
allows us to straightaway write down the degenerate in $m$ low-lying modes after substituting \eqref{TLRMP} in \eqref{CasimirRadialEqn}
\beq \label{5DSpectrum}
\frac{\Tl \omega}{\bar \kappa_+} = \frac{n\pi}{\log\left[\frac{1}{\sqrt{z_o} }\right]} 
\eeq

We will next compute the other extreme, the modes with $m_L = m_R = 0$. In this case, $\Tl \omega$ and $\omega$ are the same, so the $\omega^2$ piece drops off by \eqref{KNhorskimz} which also gives $\Lambda = l(l+2)$ from \eqref{5dAngularEquation}. The hypergeometric solution is straightforward to find and the approach proceeds parallel to the $m=0$ Kerr case discussed in Appendix B of \cite{Pradipta}. The key steps involved are making sure that the solution does not diverge\footnote{This demand can be viewed as a flatness condition or AdS normalizability, depending on one's taste in viewing the $z$ coordinate before or after the horizon-skimming limit.} at $z \rightarrow 1$ and demanding Dirichlet at $z=z_o$. This is again an analytically solvable case, and once the dust settles we find precisely \eqref{5DSpectrum}. 
The numerical calculations in the intermediate cases also show that the spectrum is degenerate with the above form, for other choices of $m$ at any given $l$ below some $l_{cut}$.

\subsection{Partition Function and Thermodynamics} \label{MPthermodynamics}

The partition function calculation proceeds parallel to that in Section \ref{KNPartitionFunction} with the only difference being that here we have an $S^3$ instead of an $S^2$. The states will be labeled by $n,l,k,m$ where $-k\leq m \leq k$, $0\leq k \leq l$ and $0\leq l \leq l_{cut}$.
For the right movers ($m>0$), the partition function integral takes the form
\bea \label{PartitionFunction5DR}
    \log Z_R &=& -\frac{N_S}{\omega'_o} \int_{0}^{l_{cut}} dl \int_{0}^{l} dk \int_{0}^{k} dm \int_{0}^{\infty} d\omega' \log(1-e^{-\beta_R \omega'})\\
   \implies Z_R(\beta_R) & = & \exp\left[\frac{\pi^2 N_S l^{3}_{cut}}{36 \beta_R \omega'_o}\right]
\eea
The only difference for the left-movers is that the integral on $m$ runs from $-k$ to $0$. In the end, we write a result similar to \eqref{SLRTLRKN}:
\beq \label{SRTRMP}
S_{L,R} = \frac{\pi^2}{3} \left(\frac{N_S l^{3}_{cut}}{6\omega'_o}\right)T_{L,R}
\eeq
which determines the left and right entropy in terms of the left and right temperature defined in \eqref{TLRMP}. Identifying the normal mode central charge with the central charge of the black hole computed in\footnote{The central charge in \cite{CK} is $6 J_\psi$. This follows the Lu-Mei-Pope (LMP) \cite{LMP} notation which sets $G^{(5)}_N = 1$. The choice of $G^{(5)}_N$ in LMP is related to that in Cvetic-Larsen (CL) \cite{CveticLarsen} via $\left(G^{(5)}_N\right)_{CL} = \frac{\pi}{4}\left(G^{(5)}_N\right)_{LMP}$. The central charge \eqref{centralcharge5D} is adapted to the CV notation. If we choose $G^{(5)}_N = \pi/4$ here, we get the central charge as $6 J_\psi$.} \cite{CK},
\beq \label{centralcharge5D}
c_L = c_R = \frac{N_S l^3_{cut}}{6\omega'_o} =  \frac{3 \pi J_\psi}{2 G^{(5)}_N}
\eeq
reproduces the correct left and right entropy:
\bea \label{SLRSimpleMP}
S_{L,R} = \frac{\pi^2 J_\psi}{2 G^{(5)}_N} \left(\frac{\bar \kappa_+ (\bar \kappa_- \pm \bar \kappa_+)}{(\bar \kappa_- - \bar \kappa_+)\bar\Omega_R - (\bar \kappa_- + \bar \kappa_+)\bar\Omega_L}\right).
\eea
These expressions can be converted to more familiar forms for the left/right/total entropies using formulas in \cite{LMP, CK, CveticLarsen}. 


\subsection{General Cvetic-Youm}

We are not aware of a Kerr-CFT calculation of the central charge and entropy of the general Cvetic-Youm black hole (with all charges and rotations turned on). In this subsection, we present a version of the calculation that seems quite natural from our holomorphically factorized normal mode perspective. A related conclusion (which is hinted at by the results in \cite{LMP, CK}) is that the calculation works with arbitrary choice for the  azimuthal direction, see Appendix \ref{azim}.

We consider the azimuthal direction defined by
\bea \label{Constraint}
    & m_L J_L + m_R J_R = 0 
\eea
We will call  $m\equiv m_R$ for convenience.  Defining $\Omega^+ \equiv \Omega_R - \frac{J_R}{J_L} \Omega_L$ and using the relation\footnote{This relation is easy to check from \eqref{OmegaLR} and \eqref{kappaplusminus}, but the fact that it holds is surprising. Its existence allows us to make a more direct connection between the Cvetic-Youm and BTZ wave equations, than say between Kerr-Newman and BTZ. (The match with BTZ is a partial consequence of the choice \eqref{Constraint} as well.)}
\beq \label{constraint5D}
\bar{\kappa}_- = \bar{\kappa}_+ \frac{J_L \Omega_R + J_R \Omega_L}{J_L \Omega_R - J_R \Omega_L} 
\eeq 
we can re-write \eqref{WaveEqn5Dz} as
\begin{align} \label{5DradialeqnCY}
&\frac{d}{d z} \left[z \frac{d}{d z} R(z)\right]+\frac{1}{4}\left[\frac{1}{z(1-z)}\left(\frac{\omega}{{\bar \kappa}_+} - \frac{m \Omega^+}{{\bar \kappa}_+}\right)^2 \right. \nonumber\\
&\left. - \frac{1}{1-z}\left(\frac{\omega}{\Bar{\kappa}_-} - \frac{m \bar \kappa_- \Omega^+}{{\bar \kappa}^{2}_+}\right)^2 + \frac{1}{(1-z)^2}\left(\left(\frac{1}{2}\left(\frac{1+z}{1-z}\right)\Delta + M\right)\omega^2 - \Lambda \right)\right] R(z) = 0 
\end{align}
after using the constraint \eqref{Constraint}. The condition \eqref{Specialmodes5D} turns into
\beq 
m^2 \left(\frac{\Delta}{2} + M\right)(\Omega^+)^2 \sim \Lambda
\eeq
and allows us to get rid of the parenthesis multiplying $1/(1-z)^2$.  The radial wave equation in \eqref{5DradialeqnCY} then reduces to the horizon-skimming wave equation. Identifying
\bea \label{CanonicalTLRCY}
&T_{L,R} = \frac{\bar \kappa^2_+}{2 \pi \Omega^+ (\bar \kappa_- \mp \kappa_+)}\\
&K_{L,R} = \mp \frac{\bar \kappa_+ \bar \kappa_-}{2 (\bar \kappa_- \mp \bar \kappa_+)}
\eea
leads to degenerate in $m$ low-lying modes analogous to \eqref{5DSpectrum}. We will not write the explicit forms here, but it should be clear that the normal modes work as we have outlined before. In this case, the Kerr-CFT central charge has not been computed from asymptotic symmetry. But it is easy to see that if we choose
\beq
c_L = c_R = \frac{N_S l^{3}_{cut}}{6\omega'_o} = \frac{6 \pi J_R}{G^{(5)}_N}
\eeq
we find the left-right entropies to be
\beq \label{SLRMP}
S_{L,R} = \frac{\pi^2 J_R}{G^{(5)}_N}\left(\frac{\bar \kappa^2_+}{\Omega^+ (\bar \kappa_- \mp \bar \kappa_+)}\right)
\eeq
Using the definition of $\Omega^+$ and \eqref{constraint5D} we can re-write it as
\beq \label{SLRSimpleCV}
S_{L,R} = \frac{2 \pi^{2}}{4 G^{(5)}_N} \frac{J_{L,R}}{\Omega_{L,R}} \bar \kappa_+
\eeq
It can be checked that this can be brought to the form of equations (7) and (8) in \cite{CveticLarsen}\footnote{We use $\beta_H = 2 \pi /\bar \kappa_+$ in \eqref{OmegaLR}. This immediately allows us to write the square roots in \cite{CveticLarsen} in terms of $J_{L,R}$, $\Omega_{L,R}$, and $\bar \kappa_+$.}. We have decided to present the above formulas for the left and right areas explicitly because they are surprisingly simple considering the fact that they apply for the general 6-charge Cvetic-Youm black hole.

The central charge result has the fairly typical form one sees in Kerr-CFT examples, and it will be interesting to derive it from an asymptotic symmetry calculation that generalizes \cite{LMP} in a suitable extremal limit.

\section{Discussion}

We have shown in this paper, that a variant of the 't Hooftian thermal atmosphere can be made precisely consistent with the thermodynamics of very general classes of black holes. A crucial ingredient was provided by the natural holomorphic factorization that emerges in the wave equations of large classes of black holes. The original calculation of 't Hooft \cite{tHooft}  was limited to Schwarzschild and had to assume both temperature and energy as inputs (instead of just say temperature, to define the ensemble), and produced only the area-scaling and not the precise pre-factor\footnote{A variant of the calculation in \cite{tHooft} is to explicitly compute the near-horizon normal modes of Schwarzschild, and work out the thermodynamics. Even if we allow free UV data ($N_S, l_{cut}$, and location of stretched horizon) this leads to an $\mathcal{O}(1)$ factor mismatch in either the entropy or the temperature for a given black hole mass. We view this as a suggestion that Schwarzschild must be viewed as a non-rotating limit of the (holomorphically factorized) Kerr black hole, if we want to think of it as a 't Hooftian quantum gas. This is again consistent with the Kerr-CFT perspective on Schwarzschild \cite{CMS}.}. 

Let us emphasize a few observations that are implicit in our calculation.
\begin{itemize}
\item Both the wave equations as well as the entropies of general classes of black holes have a holomorphically factorized structure. If the wave equations did not have such a structure, we would have no hope of connecting the two using a normal mode calculation. Previous efforts have tried to argue that the approximate low-energy symmetries of the wave equation are {\em related} to its thermodynamics, we have instead adopted the perspective that the normal modes of these wave equations are {\em responsible} for the thermodynamics. It would be very interesting to view these wave equations as low-lying effective modes of suitably defined open strings.
\item The horizon-skimming limit was crucial for identifying the quasi-degenerate set of normal modes that can carry the entropy. The near-region limit of \cite{CMS} does not lead to such a degenerate set of low-lying states. Notably,  the horizon-skimming limit still retained enough of the structures of the near-region limit, that it was compatible with Kerr-CFT. Our picture suggests that there exists an effective statistical mechanics one can associate quite universally to black holes, even when the full microphysics is not known. (Note that in all these discussions the energies are measured asymptotically.)
\item There are (at least) two perspectives that have been proposed as possible pathways to connect non-extremal black holes to string theory. One is the $c=6$ string gas suggestion of Cvetic and Larsen (see e.g., \cite{Larsenc=6}) and the other is the Kerr-CFT correspondence \cite{Guica, CMS}. The two have many common ingredients, but the precise connection has never been very clear. The $c=6$ string gas captures the level-matched structure of the effective string, but the hidden conformal symmetry is more transparent in the Kerr-CFT approach. Our discussion here seems to fall more naturally in-line with Kerr-CFT. There are hints of a long string mechanism in places in our calculation -- which may be necessary to connect the two perspectives.
\item We observed that for higher dimensional black holes the central charge is related to the black hole's angular momentum. This is not the case for BTZ, which seems to be an outlier. We make some comments on this in Appendices, but it will be good to understand it better. See \cite{Jabbari} for some possibly related discussions. A feature of BTZ that seems relevant is that because of the presence of the AdS scale, the dimensionless temperature in the non-rotating limit is finite -- in Kerr on the other hand, it diverges. This is related to our previous comment about the necessity of seeing Schwarzschild as a limit of a holomorphically factorized quantum gas.
\end{itemize}

There have been many calculations in the literature which argue that black hole entropy can be universally produced from a near-horizon 2D CFT (see e.g., \cite{Carlip}). These calculations are generally not very explicit about the definition of the CFT. Our goals in this paper were somewhat tangential to this general story, at least at the outset. We wanted to show that the 't Hooftian normal mode quantum gas (for waves around black holes) can reproduce the detailed thermodynamics of general black holes. The CFT was incidental in some ways for us, but the holomorphically factorized structure of the wave equations naturally lead us there. The horizon-skimming limit allowed us to identify the quasi-degenerate normal modes as the explicit CFT responsible for black hole entropy. It enabled us to show that making the 't Hooftian calculation work, is tantamount to fixing the normal mode central charge to be that of Kerr-CFT. We found that the relevant central charge is connected to an angular momentum mode cut-off, suggestive of a long string living in a highly twisted sector captured by the angular momentum of the black hole. It would be interesting to make this last statement precise.

\section{Acknowledgments}

We thank Vaibhav Burman for discussions. CK thanks IIT Gandhinagar for hospitality during some stages of this work. 

\appendix

\section {Thermodynamics with ``Canonical" Spectrum} \label{Canonical}

In this appendix, we will do the partition function computation of Section \ref{CasimirPartitionFunction} but with the spectrum defined with respect to the left-moving and right-moving energies conjugate to the temperatures. We define this spectrum using $H_0$ and $\bar{H}_0$ which are the Casimir generators of the $SL(2,\mathbb R)_R \times SL(2,\mathbb R)_L$ algebra for right and left modes, respectively \cite{CMS}. They are naturally conjugate to the dimensionless temperatures $T_L$ and $T_R$ as captured by the $SL(2,\mathbb R)_R \times SL(2,\mathbb R)_L$ group element \cite{CMS}
\beq 
e^{-i 4 \pi^2 T_R H_0 - i 4 \pi^2 T_L \bar{H}_0}.
\eeq

$H_0$ and $\bar H_0$ are defined in $(4.3)$ and $(4.4)$ of \cite{CMS}. A useful first step is to write them in terms of ($t,z,\phi$) using \eqref{ConfCoordRotBTZ} 
\bea \label{H0}
&H_0 = \frac{i \ T_L}{2 (K_R T_L - K_L T_R)} \partial_t - \frac{i \ K_L}{2 \pi (K_R T_L - K_L T_R)} \partial_\phi \\ \label{barH0} 
&\bar{H}_0 = -\frac{i \ T_R}{2 (K_R T_L - K_L T_R)} \partial_t + \frac{i \ K_R}{2 \pi (K_R T_L - K_L T_R)} \partial_\phi
\eea
Using the ansatz $e^{-i \omega t + i m \phi} R(z)$, the eigenvalues of $H_0$ and $\bar H_0$ is 
\bea \label{omegaR}
&\omega_R \equiv h_0 = \frac{T_L \omega}{2 (K_R T_L - K_L T_R)} - \frac{K_L m}{2 \pi (K_R T_L - K_L T_R)} \\ \label{omegaL}
&\omega_L \equiv \bar h_0 = -\frac{T_R \omega}{2 (K_R T_L - K_L T_R)} - \frac{K_R m}{2 \pi (K_R T_L - K_L T_R)}
\eea
where the first equality in both equations defines $\omega_R$ and $\omega_L$, the right and left moving spectra. 

We can now write $\omega$ in terms of $\Tl \omega$ using \eqref{tildeomegagen} and define $\Tl \omega_o$ from $\Tl \omega$ given by \eqref{CasimirSpectrum} 
\beq \label{gentilomeganot}
\Tl \omega \approx \frac{2 n \pi (K_R T_L - K_L T_R)}{(T_L + T_R) \ {\rm log}\left(\frac{1}{\sqrt{z_o}}\right)} \equiv n \Tl \omega_o
\eeq
to write $\omega_R$ and $\omega_L$ as
\beq \label{omegaRLcanonicalspec}
\omega_{R,L} = a_{R,L} n + b_{R,L} m
\eeq
where $a_{R,L}$ and $b_{R,L}$ are given as
\beq \label{aRLbRL}
a_{R,L} = \pm \frac{T_{L,R} \ \Tl \omega_o}{2 (K_R T_L - K_L T_R)} \ , \ b_{R,L} = -\frac{1}{2\pi (T_L + T_R)}
\eeq
The partition function for the right movers ($\omega_R$) with $N_S$ species of scalar fields is simply
\beq \label{PartitionFunctionomegaRnm}
\log \mathcal Z_R = -N_S \int dm \int dn \log(1-e^{-\beta_R (a_R n + b_R m)})
\eeq
Since $\omega_R$ now has a dependence on both $n$ and $m$, the $n$ and $m$ integrals do not factorize as in Section \ref{CasimirPartitionFunction}. But, we can always go in a direction orthogonal to $\omega_R$ in the $n$-$m$ plane, which we call $\omega^\perp_R$. Then, the spectrum $\omega_R$ is degenerate in $\omega^\perp_R$. Therefore, the integral in \eqref{PartitionFunctionomegaRnm} factorizes in the $\omega_R$-$\omega^\perp_R$ plane, where $\omega^\perp_R$ is 
\beq
\omega^\perp_R = -b_R n + a_R m
\eeq
Then, the partition function is 
\bea 
\log \mathcal Z_R &=& -N_S \ |{\rm det} \ \mathcal {J}_R| \int d\omega^\perp_R \int_{0}^\infty d\omega_R \log(1-e^{-\beta_R \omega_R}) \\ \label{PartitionFunctionomegaRomegaRperp}
\implies \mathcal Z_R &=& {\rm exp} \left[\frac{\pi^2 N_S m_R |{\rm det} \ \mathcal{J}_R|}{6 \beta_R}\right]
\eea
where ${\rm det} \ \mathcal J_R$ is the determinant of the jacobian transformation matrix from $n$-$m$ plane to $\omega_R$-$\omega^\perp_R$ plane, given as
\beq
{\rm det} \ \mathcal J_R = \frac{1}{a^2_R + b^2_R}.
\eeq
Let us emphasize a couple of points about the integration. Firstly, since $\beta_R$ is positive, $\omega_R$ must also be positive for the integral to converge. Hence, the integral on $\omega_R$ is from $0$ to $\infty$. This is a new set of modes from what we discussed in the main text, with their own quantization condition. Secondly, the $\omega^\perp_R$ integral must have support only on some finite range, which we call $m_R$, so that $\mathcal Z_R$ is finite. This is the analogue of $m_{cut}$ in the main text, but in the direction orthogonal to $\omega_R$.

The steps for calculating the temperature, density of states, and entropy proceed exactly identical to that in Section \ref{CasimirPartitionFunction}, and we get similar-looking expressions as in \eqref{betaRCasimir}, \eqref{SaddlePotRCasimir} and \eqref{SRCasimir}. The end result is similar to \eqref{SRTRCasimir}:
\beq \label{SRTRomegaR}
S_{L,R} = \frac{\pi^2}{3} \left(N_S m_R |{\rm det} \ \mathcal{J}_R|\right) T_{L,R}
\eeq
from which we can read off the left and right central charge as
\beq \label{centralchargeomegaLR}
c_{L,R}= N_S m_{L,R} |{\rm det} \ \mathcal{J}_{L,R}|
\eeq
Since $c_L = c_R$ in any black hole geometry (or to avoid a gravitational anomaly), we have the matching condition 
\beq 
m_L |{\rm det} \ \mathcal J_L| = m_R |{\rm det} \ \mathcal J_R| 
\eeq
Such an assumption was implicitly present in Section \ref{BTZNormalModes} as well, because we demanded $m_{cut}$ to be the same on both the left and right.

\section{Chiral Low Energy Limits: A Connection with Strings?}\label{holo}

In this appendix, we will adopt a different perspective on the modes $\omega_R$ and $\omega_L$ defined in \eqref{omegaR} and \eqref{omegaL}. The key idea is to explore the possibility that we can impose low-energy limits independently on the left and right moving modes. We will phrase the following discussion in the language of Kerr-Newman. But except for  the change in the spherical Laplacian eigenvalue depending on dimension (which results in only very minor differences), the higher dimensional cases are identical\footnote{BTZ simply amounts to setting $l=0$ on the RHS of \eqref{omegaLradialeq}.}. 

To study chiral low energy limits, we first replace $\omega$ and $m$ in terms of $\omega_R$ and $\omega_L$ in \eqref{KNangular} and \eqref{KNradialz}. The result is:
\beq \label{omegaRomegaLradialeq} 
\Bigg(\frac{d}{d z} \Big(z \frac{d}{d z}\Big)+ \frac{1}{4}\left[\frac{\left(\omega_R - \omega_L\right)^2}{z(1-z)}- \frac{\left(\omega_R + \omega_L\right)^2}{1-z}\right] + \frac{4 \ P^2}{(1-z)^2} (K_R \omega_R + K_L \omega_L)^2 \Bigg)R(z) = \frac{\kappa_l}{(1-z)^2}R(z) 
\eeq
where $P^2$ is defined as 
\beq 
P^2 \equiv \Bigg(\Big(\frac{\Tl{r}_+ - \Tl{r}_- z}{1-z}\Big)^2 + (\Tl{r}_{+}+\Tl{r}_{-})\Bigg(\Big(\frac{\Tl{r}_+ - \Tl{r}_- z}{1-z}\Big)+\Tl{r}_{+}+\Tl{r}_{-}\Bigg)-Q^2\Bigg) 
\eeq
The remarkable features we observe below arise when we set one of the modes (we will choose it to be $\omega_R$) to zero, i.e., we put them in the ground state. The motivation behind this is that for BPS black holes in string theory, putting the left or right modes in the ground state often leads to preserving some of the supersymmetry. This has led to considerable understanding of black hole physics starting with the work of \cite{Sen}. Since we have a holomorphically factorized set up (even though no supersymmetry), we can ask -- are there any insights we can gain by setting one chiral half of the oscillators in the ground state? We will see some intriguing features, but we will leave a more thorough investigation for the future.

We look at the low energy excitations of the other mode, i.e., $\omega_L \rightarrow 0$, after setting $\omega_R=0$. Since $\omega = 2(K_R \omega_R + K_L \omega_L)$ from \eqref{omegaR} and \eqref{omegaL}, the modes we consider ensures that $\omega^{2}$ piece will drop off in \eqref{KNangular} and also in \eqref{KNradialz}. This implies that $\kappa_l = l(l+1)$. Hence the radial equation that we should solve for $\omega_L$ simplifies dramatically:
\beq \label{omegaLradialeq}
\Bigg(\frac{d}{d z} \Big(z \frac{d}{d z}\Big)+ \frac{1}{4z}\omega_L^2\Bigg)R(z) = \frac{l(l+1)}{(1-z)^2}R(z). 
\eeq
The solution of \eqref{omegaLradialeq}  is 
\bea \nonumber
&R(z) = (1-z)^{l+1} \left(C_2 z^{\frac{i \omega_L}{2}} e^{-\pi  \omega_L} {}_{2}F_{1}\left(l+1,1+l+i \omega_L+1,1+i \omega _L,z\right)+ \right. \\ 
&\left. + C_1 z^{-\frac{i \omega_L}{2}} {}_{2}F_{1}\left(l+1,1+l-i
\omega_L,1-i \omega_L,z\right)\right)
\eea
The solution proceeds parallel to Section \ref{KNNormalModes}. The key steps involved are making sure that the solution does not diverge at $z\rightarrow1$ and demanding Dirichlet at $z=z_o$. This leads to the simplest of all our phase equations:
\beq
\cos\Bigg({\rm Arg}\left[\Gamma\left(1+l-i\omega_L\right)\right] + {\rm Arg}\left[\Gamma\left(i\omega_L\right)\right] -\omega_L {\rm log}\left(\sqrt{z_o}\right)\Bigg) = 0
\eeq
The normal modes can be determined analytically by using analytic approximations for arguments of the Gamma functions in the low-lying regime of $\omega_L$. The philosophy remains the same as \cite{Pradipta} and the result is:
\beq \label{lowomegaL}
\omega_L = \frac{n\pi}{\log\left(\frac{1}{\sqrt{z_o} \ l}\right)}
\eeq
where $n\in \mathbb Z^{+}$. We have retained the leading $l$-dependence because it is quite simple here. Similarly, the low-lying modes for $\omega_R$ can be obtained by setting $\omega_L = 0$ and the analytic expression remains exactly the same as \eqref{lowomegaL}. We have worked with Kerr-Newman here, but 5D black holes also lead to identical results, modulo the caveat that one needs to replace $l$ with $\frac{\sqrt{1+4l(l+2)}-1}{2}$. This is because the only difference (for any 5d black hole) is that the $l(l+1)$ on the RHS of \eqref{omegaLradialeq} gets replaced by $l(l+2)$.  

The radial equation for BTZ, once we set $\omega_R=0$ (without any further demands on $\omega_L$) directly takes the form \eqref{omegaLradialeq} with $l=0$:
\beq \label{omegaLradialeqBTZ}
\Bigg(\frac{d}{d z} \Big(z \frac{d}{d z}\Big)+ \frac{1}{4z}\omega_L^2\Bigg)R(z) = 0
\eeq
It has the solution
\beq \label{omegaLradialsolBTZ}
R(z) = C_1 \cos(\omega_L \log \sqrt{z}) + C_2 \sin(\omega_L \log \sqrt{z})
\eeq 
Demanding vanishing boundary condition at the boundary i.e., $z\rightarrow1$ leads to $C_1 = 0$ and demanding Dirichlet boundary condition at $z=z_o$ leads to the {\em exact} analytic spectrum for $\omega_L$ as:
\beq \label{omegaLBTZ}
\omega_L = \frac{n\pi}{\log\left(\frac{1}{\sqrt{z_o}}\right)}
\eeq 
where $n \in \mathbb Z^{+}$. Setting $\omega_L = 0$ leads to exactly analogous statements for $\omega_R$. 

The philosophy we have outlined in this Appendix is intriguing in its universality, simplicity and in its ability to distinguish between BTZ and higher dimensional black holes. But it is fundamentally unsatisfactory. Setting $\omega_R=0$ implies (if we take it literally) that the degeneracy is drastically reduced in the system because it uniquely fixes an $m$ once we specify an $\omega_L$. But we are hopeful that this holomorphically factorized structure, together with some previously explored ideas like Hagedorn effects at the stretched horizon \cite{Sen} or long strings \cite{Das, SusskindFat} can lead to a more concrete proposal for a stringy picture of the quantum horizon.

\section{Comments on $m \sim l$ Modes}\label{kappaextreme}

The coefficient of $m^2$ in \eqref{SpecialmodesKN} (which we will call $\alpha$ in this appendix) is superficially a function of $\tilde r_+, \tilde r_-$ and $Q$, but by dividing out suitably by (say) $Q^2$ we can turn it into a function of two variables 
\bea
\alpha = \alpha(x,y), \ \ {\rm where} \ \ x \equiv \frac{r_+}{Q}, \  \ y \equiv \frac{r_-}{Q}.
\eea
The function $\alpha$ is somewhat ugly, but remarkably, it is possible to check that it lies in the range $0$ and $1.75$, with the maximum attained as $y \rightarrow x$, which corresponds to extremality.

Note that there is a bit of an ambiguity about which modes to keep, because of our ignorance of the values of $\kappa_l$ away from spherical symmetry. This is because if we blindly assume that the value of $\kappa_l$ is $l(l+1)$ even away from spherical symmetry, then for $m \sim l$, it is clear that $ (m^2  \times 1.75)$ can overwhelm the $-l(l+1)$ in \eqref{SpecialmodesKN}. Such modes are reminiscent of modes that are not trapped by the angular momentum barrier\footnote{Modes trapped by the angular momentum barrier were the ones retained by 't Hooft in his \cite{tHooft} calculation. Our horizon-skimming limit can be viewed as a precise way to implement it for rotating black holes.}, so presumably one should exclude them. This simply changes our definition of the normal mode central charge \eqref{SLRTLRKN} by an $\mathcal{O}(1)$ factor and can be absorbed into the one parameter freedom. 

But there exists a more interesting possibility. Note that $\kappa_l$ is the Casimir eigenvalue of the ``spherical harmonic" type equation in the Kerr-Newman black hole. Its values are only known numerically away from spherical symmetry -- the cases that we are aware of where they have been explicitly computed are close to the spherically symmetric case \cite{Teukolsky}, where $\kappa_l \sim l(l+1)$. It is a possibility that the $\kappa_l$ becomes somewhat larger as the black hole starts rotating, reaching a maximum that is sufficient to balance the maximum of $\alpha$ in the extremal limit. This would mean that we do not have to exclude any modes that lie below $l_{cut}$. We have checked that for the uncharged limit of the Kerr-Newman black hole also the bound in $\alpha$ remains 1.75. For the non-rotating case, $\alpha \equiv 0$. Both these indicate that the rotation of the black hole is behind the non-vanishing nature of $\alpha$, again raising the possibility that a better understanding of $\kappa_l$ may ameliorate the need to exclude any modes.

\section{Comments on the Analytic Low-Lying Spectrum}\label{AnalyticSpec}

It is possible to do a slightly more sophisticated analytic approximation (what was called ``analytic low-lying spectrum" in \cite{Pradipta}) than our degenerate approximation, so that we retain the leading dependence on $l$ at low $l$. This is a slightly better fit to the exact (numerical) spectrum at low $l$ than the degenerate approximation we have used, but it breaks down completely and diverges at $l_{max} \sim \frac{1}{\sqrt{z_o}}$. An observation that was made regarding BTZ in \cite{Vaibhav, Pradipta} was that the value of $m_{cut}$ seems to always lie below $m_{max}$ (the analogue of $l_{max}$ for BTZ) where the analytic low-lying spectrum undergoes complete breakdown\footnote{Note that the exact numerical spectrum is more degenerate than the analytic low-lying spectrum at higher $l$, where the latter starts departing from the exact result. So the degeneracy approximation may be a better approximation than the analytic low-lying spectrum approximation at relatively larger $l$. Since we only need values of $l$ below $l_{cut}$, we will not discuss this.}. In other words, $m_{cut}$ is hierarchically lower by a factor of $\log N$ than $m_{max}$. We will see that similar statements remain true even in higher dimensions. The way this works out relies on the fact that the degeneracy of states at a given $l$ scales differently in different dimensions. We discuss this briefly below for the Kerr-Newman case.

Let us start by noting that we can solve for $l^{2}_{cut}$ from \eqref{centralchargeKN} as 
\beq \label{lcutKN}
l^{2}_{cut} = \frac{12 \pi (\Tl r_+ + \Tl r_-) \sqrt{\Tl r_+ \Tl r_- - Q^2}}{N_S l^{2}_{p} \log\left(\frac{1}{\sqrt{z_o}}\right)}.
\eeq
We have used our definition of $J$ from above \eqref{KNangular} and also set $G^{(4)}_N = l^{2}_p$ in $3+1 \ {\rm D}$. Now, note that $l^{2}_{max} \sim \frac{1}{z_o}$ is the location where the analytic low-lying spectrum breaks down, as observed above\footnote{We do not need detailed expressions for the analytic low-lying spectrum. The only fact we will need is that it breaks down at $l \sim 1/\sqrt{z_o}$. The interested reader can take a look at related expressions in \cite{Pradipta, Vaibhav}.}. This means that
\beq \label{lcutlmaxKN}
\frac{l^{2}_{cut}}{l^{2}_{max}} \sim \frac{ z_o (\Tl r_+ + \Tl r_-)\sqrt{\Tl r_+ \Tl r_- - Q^2}}{N_S l^{2}_p \log \left(\frac{1}{\sqrt{z_o}}\right)}
\eeq
If we fix $z_o$ to be at about a Planck length from the horizon by setting $z_o \sim \frac{l^{2}_p}{4 (r^{2}_+ + a^2)}$ in \eqref{lcutlmaxKN}, we see that there is a small hierarchy between $l^{2}_{cut}$ and $l^{2}_{max}$ which is provided by the $\sim \log (1/l_p) \sim \log N$ in the denominator\footnote{The terms dependant on black hole parameters are $\mathcal O(1)$.}. This hierarchy gets larger as the number of species ($N_S$) increases. 

What the above calculation suggests is that even in higher dimensions, the thermodynamics of black holes seems to be supported by a regime  of normal modes below $l_{max}$. This observation is crucially reliant on the dimensionality of spacetime on both the right and left hand sides of \eqref{centralchargeKN}. On the right hand side, the scaling emerges because of the length scaling of Planck's constant in 4 dimensions. On the left hand side, precisely the same scaling emerges for an apparently different reason -- namely that the number of angular momentum modes in $4$ dimensions scales as $l_{cut}^2$. This correspondence holds in higher dimensions as well. 

To see this slightly more explicitly, lets us consider the possibility that \eqref{centralchargeKN} had just one power of $l_{cut}$ instead of $l^{2}_{cut}$. This was the case in 2+1 dimensions and it would imply that
\beq
l_{cut} \sim \frac{1}{N_S l^{2}_p \log\left(\frac{1}{\sqrt{z_o}}\right)}
\eeq
On the other hand $l_{max}$ always scales as $1/\sqrt{z_o}$. If we set $z_o = \frac{l^{2}_p}{4 (r^{2}_+ + a^2)}$ this would lead to an $l_{cut}$ that is hierarchically larger than $l_{max}$ (by approximately a factor of $\sim N/\log N$). Note that the exact spectrum is in fact still quite degenerate at large $l$ as long as $z_o$ is small enough, so at least for the moment we simply view this as a technical observation about the analytic low-lying spectrum approximation. 

\section{General Azimuthal Direction}\label{azim}

As mentioned in Section \ref{MPmodes}, our set up suggests a natural freedom in the azimuthal direction along which we decide to define our horizon-skimming limit. We explore the most general such choice in this Appendix. We show that this leads to well-defined right-left temperatures as well as a central charge that is naturally left-right matched. Our result here can be viewed as a one parameter family of Kerr-CFT correspondences for the 5D Cvetic-Youm black hole, that generalizes the discretely many cases noted in \cite{LMP, CK}.

We consider an azimuthal direction defined by two arbitrary real parameters $a$ and $b$\footnote{We have used the notations $a$ and $b$ in Section \ref{BTZNormalModes} also in a different context. The present use of $a$ and $b$ is restricted to this Appendix.}:
\beq
a m_R + b m_L = 0
\eeq
together with the understanding that $m_R\equiv m$ in the rest of this Appendix. Following the same sequence as in the main text, this leads us to
\bea \label{TLRgenazi}
& T_L = \frac{\bar \kappa_+ (\bar \kappa_- + \bar \kappa_+)}{2 \pi \Omega_R (\bar \kappa_- - \bar \kappa_+)\left(1+\frac{b \Omega_L (\bar \kappa_- + \bar \kappa_+)}{a \Omega_{R} (\bar \kappa_- - \bar \kappa_+)}\right)}\ , \ T_R = \frac{\bar \kappa_+}{2 \pi \Omega_R \left(1+\frac{b \Omega_L (\bar \kappa_- + \bar \kappa_+)}{a \Omega_{R} (\bar \kappa_- - \bar \kappa_+)}\right)} \\ \label{KLRgenazi}
& K_L = -\frac{\bar \kappa_+ \bar \kappa_-}{(\bar \kappa_- - \bar \kappa_+)\left(1+\frac{b \Omega_L (\bar \kappa_- + \bar \kappa_+)}{a \Omega_{R} (\bar \kappa_- - \bar \kappa_+)}\right)} \ , \ K_R = \frac{\bar \kappa_+ \bar \kappa_-}{(\bar \kappa_- + \bar \kappa_+) \left(1+\frac{a \Omega_{R} (\bar \kappa_- - \bar \kappa_+)}{b \Omega_L (\bar \kappa_- + \bar \kappa_+)}\right)}
\eea
Using \eqref{SLRSimpleCV} together with the definitions of $T_{L,R}$ in \eqref{SRTRMP}, leads us to the left and right central charges
\beq
c_L = c_R = \frac{N_S l^{3}_{cut}}{6 \omega'_o} = \frac{3\pi J_R}{G^{(5)}_N} \left(1+\frac{b \Omega_L (\bar \kappa_- + \bar \kappa_+)}{a \Omega_{R} (\bar \kappa_- - \bar \kappa_+)}\right)
\eeq
where for finding the $c_L$ we have used the constraint \eqref{constraint5D}. It is remarkable that we are able to get a left-right matched central charge, thanks to the constraint \eqref{constraint5D} satisfied by black hole parameters.

\section{Normal Modes of Extremal Black Holes}\label{AdS2-sec}

In this Appendix, we will show that the normal modes of extremal black holes do not have the quasi-degeneracy in $l$ that played a crucial role in our calculations. For the rotating BTZ case in the extremal limit, a related numerical observation was made in \cite{ArnabSuman}. We wish to demonstrate this as a general property of the near-horizon AdS$_2$ geometry that appears in the extremal limit. These observations may be significant for the conventional fuzzball program (as defined in \cite{Vaibhav2}). 

We first review some well-known facts to be clear about the dimensionfull quantities. With the parameters for the mass and electric charge being $M$ and $Q$ respectively, the Reissner-Nordstrom metric is 
\bea
& ds^2 = -\frac{(r-r_+)(r-r_-)}{r^2} dt^2 +\frac{r^2}{(r-r_+)(r-r_-)} dr^2 + r^2 d\Omega^2_{2} \\
& r_\pm = Q l_p + E l^{2}_p \pm \sqrt{2 E Q l^{3}_p + E^{2} l^{4}_p}
\eea
where $r_+$ and $r_-$ are the outer and inner horizons.
\beq
E = M - \frac{Q}{l_p}
\eeq
is the excitation energy above extremality and the Planck length, $l_p = \sqrt{G_N}$. In the extremal case $E=0$, only $l_p$ carries dimension. So, this is the parameter that defines what ``near” means in a near-horizon limit. Thus, we can define a new coordinate (see e.g., \cite{GaborSarosi})
\beq
z = \frac{Q^2 l^2_{p}}{r-r_+}
\eeq
and zoom into $r_+$ by taking $l_p \rightarrow 0$ while holding $z$ fixed. The resulting metric is 
\beq \label{AdS2Poincare}
ds^2 \approx l^{2}_p Q^2 \left(\frac{-dt^2 + dz^2}{z^2} + d\Omega^{2}_2\right)
\eeq
which is the product space ${\rm AdS}_2 \times S^2$ with the ${\rm AdS_2}$ part in Poincare coordinates. $l^2_p Q^2$ can be interpreted as the ${\rm AdS}_2$ length scale. 

We will find the normal modes in this AdS$_2$ geometry with a cut-off close to the Poincare horizon. For a massless scalar $\Phi(t,r,\phi) = e^{-i\omega t + i m \phi} R(z) S(\theta)$ the radial equation in the near horizon background \eqref{AdS2Poincare} is:
\bea 
& z^2 \left(\frac{d^2 R(z)}{d z^2} +\omega^2 R(z)\right) = l(l+1) R(z) \label{extremalRNradial}.
\eea
The general solution has the form
\beq \label{RadialSolnExtremalRN}
R(z) = \sqrt{z} \left(C_1 {\rm J}\left[l+\frac{1}{2}, z \omega\right]+C_2 {\rm Y}\left[l+\frac{1}{2}, z \omega\right]\right)
\eeq
where $\rm J$ and $\rm Y$ stand for Bessel functions of $1^{\rm st}$ and $2^{\rm nd}$ kind respectively. Near the boundary $z \rightarrow 0$  \eqref{RadialSolnExtremalRN} looks like
\bea 
\omega^{-\frac{1}{2}-l}\left(-\frac{C_2 2^{l+\frac{1}{2}} z^{-l} \Gamma \left[l+\frac{1}{2}\right]}{\pi} + \frac{2^{-l-\frac{1}{2}} \omega ^{3 l} z^{l+1} (C_1-C_2 \tan (\pi  l))}{\Gamma \left(l+\frac{3}{2}\right)}\right)
\eea
Fluctuations in AdS$_2$ cause too much backreaction, but let us be liberal\footnote{Our point here is that despite this, we will not find the quasi-degenerate spectrum.} and allow modes that satisfy ``regularity" conditions near the boundary, $z \rightarrow 0$. Working with $l>0$, we see that the first term in the parenthesis is divergent. So, our regularity demand will be $C_2 = 0$. By doing a near horizon expansion ($z \rightarrow \infty$) of the remaining solution we get, 
\beq \label{extremalRNRadialNH}
C_1 \left (\frac{\sqrt{\frac{2}{\pi }} \sin \left(\omega 
   z - \frac{\pi  l}{2}\right)}{\sqrt{\omega }} + \frac{l (l+1) \cos \left(\omega  z - \frac{\pi  l}{2}\right)}{\sqrt{2 \pi } \omega ^{3/2} z} + \mathcal{O}\left(\frac{1}{z^2}\right)\right)
\eeq
Now demand a Dirichlet boundary condition at the stretched horizon radius, $z = z_o$. The phase equation
\beq
\sin \left(\omega z_o - \frac{\pi l}{2}\right) = 0
\eeq
gives the spectrum as
\beq
\omega = \frac{(2 n + l) \pi}{2 z_o}
\eeq
where, $n,l \in \mathbb{Z^+}$.
This is linear in $l$, as familiar form higher dimensional {\em global} AdS. There is no exponentially quasi-degeneracy.

\end{document}